\documentclass[prb,twocolumn,showpacs]{revtex4}
\usepackage{graphicx}
\usepackage{bm}
\usepackage{longtable}

\begin{document}

\title{Dynamic Simulation of Structural Phase Transitions in Magnetic Iron}
\author{Pui-Wai Ma}
\email{Leo.Ma@ukaea.uk}
\affiliation{Culham Centre for Fusion Energy, Abingdon, Oxfordshire OX14 3DB, United Kingdom}
\author{S. L. Dudarev}
\affiliation{Culham Centre for Fusion Energy, Abingdon, Oxfordshire OX14 3DB, United Kingdom}
\author{Jan S. Wr\'obel}
\affiliation{Division of Materials Design, Faculty of Materials Science and Engineering, Warsaw University of Technology, Wo{\l}oska 141, 02-507 Warsaw, Poland}

\begin{abstract}
The occurrence of bcc-fcc ($\alpha$-$\gamma$) and fcc-bcc ($\gamma$-$\delta$) phase transitions in magnetic iron stems from the interplay between magnetic excitations and lattice vibrations. However, this fact has never been proven by a direct dynamic simulation, treating non-collinear magnetic fluctuations and dynamics of atoms, and their coupling at a finite temperature. Starting from a large set of data generated by \textit{ab initio} simulations, we derive non-collinear magnetic many-body potentials for bcc and fcc iron describing fluctuations in the vicinity of near perfect lattice positions. We then use spin-lattice dynamics simulations to evaluate the difference between free energies of bcc and fcc phases, assessing their relative stability within a unified dynamic picture. We find two intersections between the bcc and fcc free energy curves, which correspond to $\alpha$-$\gamma$ bcc-fcc and $\gamma$-$\delta$ fcc-bcc phase transitions. The maximum fcc-bcc free energy difference over the temperature interval between the two phase transition points is 2 meV, in agreement with other experimental and theoretical estimates.
\end{abstract}

\pacs{75.50.Bb, 75.10.Hk, 02.70.Ns, }

\maketitle

\section{Introduction}
Pure iron undergoes bcc-fcc ($\alpha$-$\gamma$) and fcc-bcc ($\gamma$-$\delta$) phase transitions at $T_{\alpha-\gamma}$=1185K and $T_{\gamma-\delta}$=1667K, respectively. They occur in the temperature interval between the Curie temperature $T_C=1043$K and the melting temperature $T_M$=1811K. These transitions have exceptionally significant practical implications as they are responsible for the formation of martensite in steels, and hence represent the most basic phenomena underpinning steel manufacturing and modern metallurgy. It has long been speculated that $\alpha$-$\gamma$-$\delta$ phase transitions in iron stem from the interplay between magnetic excitations and lattice vibrations. Still, there is no explicit proof, derived from a direct simulation, confirming this assertion. The position is somewhat unsatisfactory as it leaves open the fundamental question of whether, with modern materials modelling concepts and algorithms, it is actually possible to discover a magnetism-driven structural phase transition by exploring the dynamics of a discrete atomistic model.

Hasegawa and Pettifor\cite{HasegawaPRL1983} investigated the relative stability of bcc, fcc, and hcp phases of iron as a function of temperature and pressure. They concluded that the relative stability of phases was primarily determined by the magnetic free energy contribution. Since they used a single-site spin-fluctuation approximation, which is a mean-field approach where the short-range magnetic order (SRMO) is neglected, the predicted phase diagram was rather qualitative than quantitative.

Recent experiments\cite{NeuhausPRB2014}, which explored phonon dispersion in iron at high temperatures, showed that the stabilization of the high-temperature bcc $\delta$ phase is due primarily to vibrational entropy, whereas the fcc $\gamma$ phase is stabilized by the fine balance between electronic and vibrational entropy contributions. This agrees with calculations performed using Monte Carlo Magnetic Cluster Expansion (MCE) \cite{LavrentievCMS2010,LavrentievPRB2010}. Although in the MCE the phonon contribution to the free energy is derived from experimental data, the MCE analysis shows that magnetic excitations lower the free energy difference between the bcc and fcc phases $\Delta F^{fcc-bcc}$, and stabilize the $\gamma$ phase. The bcc $\delta$ phase again becomes more stable at higher temperatures, because the lattice vibrations part of the free energy at high temperature is greater than the part associated with magnetic excitations.

Several recent studies of phase stability of iron are based on \textit{ab initio} calculations\cite{HerperPRB1999,SoulairolJPCM2010,NguyenPRB2009,AlnemratJPCM2014} or the tight-binding Stoner model\cite{SoulairolJPCM2010,NguyenPRB2009,AutesJPCM2006}. Most of them explore phase stability at 0K, since this is an intrinsic limitation associated with density functional theory (DFT). Treating magnetism in the framework of a tight-binding model is also not trivial since the fully non-collinear version of the Hamiltonian, including both spin and orbital magnetism, has been derived only recently\cite{CouryPRB2016}. In principle, the Coury Hamiltonian\cite{CouryPRB2016} should enable fully self-consistent non-collinear magnetic dynamic simulations of atoms and magnetic moments, treated at the electronic scale.

Several approaches have been developed to describe finite-temperature magnetic excitations using \textit{ab initio} techniques, see Ref. \onlinecite{KormannCOSSMS2016} for a review. The disordered local moments (DLM)\cite{OkatovPRB2009,RazumovkiyPRL2011,PolesyaPRB2016,MankovskyPRB2013} approximation assumes randomly distributed collinear up and down oriented magnetic moments, to imitate a fully magnetically disordered paramagnetic (PM) state of a material. It has been applied to modeling bcc-fcc\cite{OkatovPRB2009} and bcc-hcp\cite{MankovskyPRB2013} transitions. To treat temperatures lower than $T_C$, where iron is in a partially ordered ferromagnetic (FM) state, partial or uncompensated DLM appromations\cite{RazumovkiyPRL2011,PolesyaPRB2016, MankovskyPRB2013} were proposed, where the net magnetization is constrained to a fixed value, matching experimental observations. This approach reproduces the elastic anisotropy of iron and Fe-Cr alloys\cite{RazumovkiyPRL2011}. However, the notion of SRMO remains undefined as the DLM approximation neglects magnetic non-collinearity.

Recently, Leonov {\it et al.}\cite{LeonovPRL2011,LeonovPRB2012,LeonovSR2014} investigated the structural stability of iron using a combination of DFT and dynamical mean field theory (DMFT). The treatment involved an explicit consideration of temperature-dependent electron correlations. In the DMFT formalism, excitations associated with electron-electron interactions were treated using a single-site mean-field approximation, and neither collective magnetic excitations nor the SRMO were taken into account.

SRMO can be treated in the random phase approximation (RPA) combined with rescaling \cite{KormannPRB2008}. This approach was applied to evaluate magnetic, electronic and phonon contributions to the free energy, and also to assess the pressure dependence of $T_C$ \cite{KormannPRB2009} in iron. The treatment has also been extended to quantum effects by rescaling the available classical solutions \cite{KormannPRB2010}. However, the mean-field nature of the approach gives rise to the predicted value of $T_C$ to be higher than the observed value. An alternative approach to modelling SRMO is the spin-wave method\cite{RubanPRB2012}.

SRMO can be simulated using \textit{ab initio} spin dynamics (SD)\cite{AntropovPRL1995,AntropovPRB1996,IvanovJAP1999,FahnleCMS2005} combined with constrained non-collinear calculations\cite{MaPRB2015,KurzPRB2004,SingerPRB2005, UjfalussyJAP1999,StocksPMB1998,DederichsPRL1984}. However, this is a computationally highly demanding approach, applicable only to relatively small systems. A more practical way of treating SRMO is by deriving parameters from {\it ab intio} data and performing atomic scale SD\cite{MaPRB2011,MaPRB2012} or Monte Carlo \cite{RubanPRB2007,RubanPRB2013,LavrentievCMS2010,LavrentievPRB2010} simulations.

In the treatment of structural phase transitions, magnetic and phonon excitations, as well as coupling between them, appear significant. K\"ormann {\it et al.} proposed a spin space averaging procedure \cite{KormannPRB2012,KormannPRL2014} to evaluate effective interatomic forces at a finite temperature by interpolating wave functions between the FM and PM states and performing statistical averaging over many magnetic microstates. The procedure was applied to phonon spectra and changes in the spectra due to magnon-phonon interactions.

Conventional many-body interatomic potentials for molecular dynamics (MD) treat only the atomic degrees of freedom. Many of the potentials \cite{OlssonPRB2005, OlssonPRB2006, MalerbaJNM2010,MarinicaPRL2012,LeeJPCM2012} are fitted to \textit{ab initio} data at 0K. Although some were successful\cite{LeeJPCM2012} in reproducing bcc, fcc, hcp phases at different temperatures and pressures, they required adjusting the energies of fcc and hcp structures to compensate for the absence of an explicit treatment of magnetic effects. Improvements in the functional forms of interatomic potentials were proposed\cite{DudarevJPCM2005,DerletPMS2007,AcklandJNM2006}, but they still did not include magnetic degrees of freedom explicitly, and did not treat magnetic thermal excitations.

There were partially successful attempts to incorporate magnetic states explicitly in MD. For example, Lian {\it et al.} \cite{LianPRB2015} performed \textit{ab initio} MD simulations to obtain the phonon dispersion of $\gamma$ and $\delta$ phases over a range of temperatures. However, only the antiferromagnetic state (AFM) was considered as representative of the PM state. Alling {\it et al.}\cite{AllingPRB2016} performed DLM+MD to study the effect of atomic vibrations on magnetic properties. No equations of motion for magnetic moments were considered, and the dynamics of magnetic excitations was modeled as stochastic spin flips. This leads to conceptual difficulties in the treatment of thermalization of atoms and spins.

In this study, we use spin-lattice dynamics (SLD)\cite{MaACP2008,MaPRB2008} to study structural phase transitions in magnetic iron. SLD treats the dynamics of lattice and magnetic subsystems within a unified framework. Lattice and spin temperatures\cite{MaPRE2010} can be well controlled through Langevin thermostats\cite{ChandrasekharRMP1943,KuboRPP1966,MaPRB2012}. SLD also treats anharmonicity and the coupling of lattice vibrations to magnetic excitations. SLD is an efficient and versatile simulation approach, and it has been recently applied to a variety of phenomena including the anomalous self-diffusion in iron\cite{WenJNM2013,WenJNM2014}, magnetic excitations in thin films\cite{MaPM2009} and a broad range of other magnetic phenomena \cite{BeaujouanPRB2012,ThibaudeauPASMA2012,PereraJAP2014,PereraPRB2016}.

An assessment of structural phase stability requires the evaluation of magnetic free energy, where the treatment of SRMO is critically important\cite{KormannPRL2014,RubanPRB2012} to the quantitative prediction of phase transition temperatures. The SRMO is treated fully by SLD through the use of dynamic spin equations of motion. In this work, interatomic interaction parameters for SLD simulations are derived from \textit{ab initio} calculations, and are given in the form of a non-collinear magnetic many-body potential. Using SLD simulations, we are able to evaluate contributions to the free energy from lattice and spin excitations as functions of temperature. By a direct simulation, we find $\alpha-\gamma$ and $\gamma-\delta$ phase transitions, which manifest themselves as changes of sign of $\Delta F^{fcc-bcc}$. The maximum free energy difference between bcc and fcc phases over the temperature interval between the two phase transition points is close to 2 meV. 

\section{Free Energy Calculations}
We used two complementary techniques to carry out free energy calculations, the umbrella sampling and thermodynamic integration. Both methods are well established but they have not yet been applied to the treatment of phase transitions in magnetic systems. Brief summaries of the techniques are given below. We also describe our approach to the evaluation of the free energy of a harmonic oscillator and a Landau oscillator in the classical limit, and outline our sampling procedure.

Since we only treat the classical limit, our results are valid at temperatures that are sufficiently high, close to or above approximately one-third of the characteristic temperature for a particular subsystem, the Debye temperature for lattice vibrations and the Curie or the N\'eel temperature for magnetic excitations. Low temperatures classical results are only given for completeness, and they should not be treated as predictions. A quantum treatment, not considered below, should be applied if one is interested in the accurate low temperature values.

In lattice case, the average number of phonons in mode $k$, $\langle n_k \rangle$, is given by the Planck distribution, where at sufficiently high temperatures
$$
\langle n_k\rangle +{1\over 2}={1\over {\exp \left({\hbar \omega_k \over k_BT}\right)-1}}+{1\over 2}\approx {k_BT\over \hbar \omega_k}+{1\over 12}\left({\hbar \omega_k\over k_BT}\right)+...
$$ 
Taking as an estimate $\hbar\omega _k \sim k_BT_D$, we see that classical treatment applies at $T\ge T_D/\sqrt{12}\approx T_D/3.5$. The Debye temperature of iron is close to 470K. Therefore, one can argue that the temperature range of validity of classical molecular dynamics is defined by the condition $T> 135$K. A broadly similar argument can also be applied to magnetic excitations, which we treat classically at temperatures above approximately one third of $T_C$ or $T_N$.

\subsection{Umbrella Sampling}
Umbrella sampling\cite{TorrieJCP1977} is a biased sampling technique. It is a re-weighting technique for evaluating the free energy difference between a reference and a target state. It is particularly useful for sampling metastable states.

We start by considering two classical Hamiltonians $\mathcal{H}^0$ and $\mathcal{H}^1$, and their difference:
\begin{equation}
\delta\mathcal{H}_{um}= \mathcal{H}^1 - \mathcal{H}^0.
\label{eq1}
\end{equation}
The ensemble average of an observable $\mathcal{O}$ with respect to Hamiltonian $\mathcal{H}^0$ at a particular temperature $T$ is:
\begin{equation}
\langle \mathcal{O} \rangle_{0} = \frac{\int \mathcal{O} \exp(-\beta\mathcal{H}^0) d\Omega}{\int \exp(-\beta\mathcal{H}^0) d\Omega}
\label{eq2}
\end{equation}
where $\beta=(k_B T)^{-1}$ and $d\Omega$ is an element of volume in classical phase space, which in this instance has 9N dimensions, and includes position vectors of all the atoms, their kinematic momenta, and vectors of all the atomic magnetic moments. Substituting  (\ref{eq1}) into (\ref{eq2}), we arrive at:
\begin{eqnarray}
\langle \mathcal{O} \rangle_{0} = \frac{\langle \mathcal{O} \exp(\beta \delta\mathcal{H}_{um}) \rangle_{1}}{\langle\exp(\beta \delta\mathcal{H}_{um}) \rangle_{1}}.
\label{eq3}
\end{eqnarray}
This formula recasts the calculation of an ensemble average of a classic observable $\mathcal{O}$ over the equilibrium defined by Hamiltonian $\mathcal{H}^0$ into calculations of ensemble averages of $\mathcal{O} \exp(\beta \delta\mathcal{H}_{um}) $ and $\exp(\beta \delta\mathcal{H}_{um})$ over thermodynamic equilibrium defined by another Hamiltonian $\mathcal{H}^1$.

This shows a way of evaluating the difference between free energies associated with two classical Hamiltonians $\mathcal{H}^0$ and $\mathcal{H}^1$. For example, the expression for the free energy corresponding to Hamiltonian $\mathcal{H}^0$ can be written as
\begin{eqnarray}
F^0 &=& -k_B T \ln \int \exp(-\beta\mathcal{H}^0) d\Omega\nonumber\\
&=&-k_B T \ln \int \exp(-\beta(\mathcal{H}^0-\mathcal{H}^1+\mathcal{H}^1) d\Omega\nonumber \\
&=&-k_B T \ln \int \exp(\beta\delta \mathcal{H}_{um}-\beta\mathcal{H}^1) d\Omega\nonumber \\
&=&-k_B T \ln \left\{ {\left[\int \exp(\beta\delta \mathcal{H}_{um}-\beta\mathcal{H}^1) d\Omega \over \int \exp(-\beta\mathcal{H}^1) d\Omega \right] }\right. \nonumber \\
&\times& \left.\int \exp(-\beta\mathcal{H}^1) d\Omega \right\}\nonumber \\
&=& -k_B T \ln \langle \exp (\beta\delta \mathcal{H}_{um}) \rangle_{1} + F^1.\label{umbrella}
\end{eqnarray}
Hence, the difference between free energies of two equilibrium configurations defined by Hamiltonians $\mathcal{H}^0$ and $\mathcal{H}^1$ is
\begin{eqnarray}
\delta F_{um} &=& F^1 - F^0 \nonumber \\
&=& k_B T \ln \langle \exp (\beta\delta \mathcal{H}_{um}) \rangle_{1}
\end{eqnarray}
If one of the free energies $F^1$ is known, the other free energy $F^0$ can be computed by sampling the phase space with thermodynamic weights defined by $\mathcal{H}^1$, and no independent averaging over thermodynamic equilibrium defined by $\mathcal{H}^0$ is required.

\subsection{Thermodynamic Integration}
Another technique for evaluating the difference between free energies is the adiabatic switching thermodynamic integration method \cite{KirkwoodJCM1935,FrenkelJCP1984,CiccottiBOOK1987}. For any two Hamiltonians $\mathcal{H}^0$ and $\mathcal{H}^1$, we can define a Hamiltonian that is a linear combination
\begin{equation}
\mathcal{H}_{ti}(\lambda)=(1 -\lambda)\mathcal{H}^0+\lambda\mathcal{H}^1,
\end{equation}
where $\lambda$ is a switching parameter varying from 0 to 1. The difference between Hamiltonians $\mathcal{H}^1$ and $\mathcal{H}^0$ equals the derivative of $\mathcal{H}_{ti}$ with respect to $\lambda$.
\begin{eqnarray}
\delta\mathcal{H}_{ti} &=& \mathcal{H}^1 -\mathcal{H}^0=\frac{\partial\mathcal{H}_{ti}}{\partial\lambda}
\end{eqnarray}
The free energy difference between the initial ($\lambda=0$) and final ($\lambda=1$) states can be calculated as an integral over  the switching parameter, namely
\begin{equation}
\delta F_{ti} = F^1 - F^0 = \int_0^1 \langle\delta\mathcal{H}_{ti}\rangle_\lambda d\lambda.
\end{equation}
Brackets $\langle...\rangle_\lambda$ correspond to taking an ensemble average with respect to $\mathcal{H}_{ti}(\lambda)$. We evaluate this average using a dynamic simulation, by imposing the chain rule $d\lambda = (\partial\lambda/\partial t) dt$ and adopting a time-dependent switching function \cite{KoningPRE465}:
\begin{eqnarray}
\lambda(\tau) &=& \tau^5(70\tau^4-315\tau^3+540\tau^2-420\tau+216)
\end{eqnarray}
where $\tau=t/t_{tot}$, $t$ is the elapsed time and $t_{tot}$ is the total switching time. One can check that if $t=0$ then $\lambda=0$, and if  $t=t_{tot}$ then $\lambda=1$.

\subsection{Harmonic oscillator and Landau oscillator}
We use two reference states for free energy calculations. A harmonic oscillator is used as a reference state for the free energy of the lattice. A Landau oscillator is used as a reference state for the treatment of magnetic excitations.

The Hamiltonian of a three-dimensional harmonic oscillator is
\begin{equation}
\mathcal{H}_{HO}=\frac{\mathbf{p}^2}{2m}+\frac{1}{2}m\omega^2 \mathbf{x}^2 + C,
\end{equation}
where $\mathbf{p}$ is the kinematic moment, $\mathbf{x}$ is the displacement, $\omega$ is the frequency, $m$ is the mass and $C$ is a constant. In the classical limit, the free energy can be evaluated analytically as
\begin{equation}
F_{HO}=-3k_B T \ln\left(\frac{k_B T}{\hbar\omega}\right) + C,
\end{equation}
where the Planck constant is introduced for dimensional convenience. In what follows, we assume that $\omega$ equals the Debye frequency of iron, $\hbar\omega= k_B T_D$, where $T_D=470$K.

The Landau spin Hamiltonian has the form
\begin{equation}
\mathcal{H}_{LO}=A_{LO}\mathbf{S}^2+B_{LO}\mathbf{S}^4
\end{equation}
where $A_{LO}$ and $B_{LO}$ are constants, and $\mathbf{S}$ is a dimensionless spin vector. The free energy of magnetic excitations in the Landau approximation can be written as
\begin{equation}
F_{LO}=-k_B T \ln \left(4\pi\int_0^\infty\exp (-\beta\mathcal{H}_{LO}) S^2 dS \right)
\end{equation}
where $S$ is the magnitude of $\mathbf{S}$. In this work, we choose $A_{LO}=-1.184$eV and $B_{LO}=0.578$eV to match the spectrum of longitudinal magnetic excitations\cite{MaPRB2012}. The value of $F_{LO}$ is then computed numerically at various temperatures.

Hamiltonians $\mathcal{H}_{HO}$ and $\mathcal{H}_{LO}$ defined above refer to an individual atom and an individual spin. In the calculations below we will use $\mathcal{H}_{HO}$ and $\mathcal{H}_{LO}$ to represent all the atoms and spins $N$, assuming that they are independent of each other.

\subsection{Sampling procedure - MD}
In an MD simulation, we calculate the free energy using the umbrella sampling. The full lattice Hamiltonian has the form
\begin{equation}
\mathcal{H}_l=\sum_i \frac{\mathbf{p}_i^2}{2m} + U(\mathbf{R})
\end{equation}
where $U(\mathbf{R})$ is the interatomic potential, $\mathbf{R}=\{\mathbf{R}_i\}$ are the coordinates of all the atoms, and $\mathbf{p}=\{\mathbf{p}_i\}$ are the kinematic momenta.

The free energy of the lattice system can be computed by using equation (\ref{umbrella}) and sampling over the thermodynamic equilibrium of harmonic oscillators, namely
\begin{equation}
F_l = F_{HO} - \delta F_l,
\end{equation}
where
\begin{equation}
\delta F_l = k_B T\ln\langle\exp(\beta\delta \mathcal{H}_l)\rangle_{HO}, \label{eq18}
\end{equation}
and $\delta \mathcal{H}_l$ is defined as
\begin{eqnarray}
\delta H_l &=& \mathcal{H}_{HO}-\mathcal{H}_l\nonumber \\
&=& \sum_i \left(\frac{1}{2}m\omega^2\mathbf{x}^2_i + C\right) - U(\mathbf{R}).
\end{eqnarray}
Here $\mathbf{x}_i$ is the displacement of atom $i$ from its position in the lattice $\mathbf{R}_i^0$, i.e. $\mathbf{x}_i = \mathbf{R}_i - \mathbf{R}_i^0$. The value of constant $C$ is chosen to minimize the variation of $\delta \mathcal{H}_l$. A suitable choice of $C$ helps ensure the numerical stability of umbrella sampling by eliminating large numerical values in the argument of exponential function in Eq. (\ref{eq18}).

Sampling is performed using dynamic Langevin thermostat simulatons\cite{ChandrasekharRMP1943,KuboRPP1966} that generate the correct equilibrium energy distribution, assuming ergodicity. Langevin equations of motion have the form
\begin{eqnarray}
\frac{d\mathbf{R}_i}{dt} &=& \frac{\mathbf{p}_i}{m}\nonumber \\
\frac{d\mathbf{p}_i}{dt} &=& \mathbf{F}_i -\gamma_l\frac{\mathbf{p}_i}{m}+\mathbf{f}_i,\label{eq22}
\end{eqnarray}
where the regular component of the force acting on atom $i$ is
\begin{equation}
\mathbf{F}_i=  -\frac{\partial \mathcal{H}_{HO}}{\partial \mathbf{R}_i}.
\end{equation}
The damping constant $\gamma_l$ and the fluctuating force $\mathbf{f}_i$ are related through the fluctuation-dissipation theorem\cite{ChandrasekharRMP1943,KuboRPP1966}, namely $\langle f_{\alpha i}(t)f_{\beta j}(t') \rangle = 2k_B T \gamma_l \delta_{\alpha \beta} \delta_{ij}\delta (t-t')$, where indexes $\alpha$ and $\beta$ refer to Cartesian coordinates $x,y,z$.

By following the above procedure, we sample over thermodynamic equilibrium defined by the Einstein model for a solid, where the lattice points are ordered as either bcc or fcc lattices. A major advantage of umbrella sampling is that it overcomes the difficulties associated with sampling the spectra of excitations of an unstable structure. For example, the recently developed interatomic potentials for iron\cite{MalerbaJNM2010,MarinicaPRL2012} predict a stable bcc phase. The fcc phase is unstable, but since sampling is performed over an equilibrium defined by suitably spatially ordered harmonic oscillators, the fact that the crystal structure is unstable has no effect on the stability of the numerical procedure.

\subsection{Sampling procedure - SLD}
In a spin-lattice dynamic (SLD) simulation, we adopted a two-step approach to free energy calculations. We use the umbrella sampling, which is followed by thermodynamic integration. We write the spin-lattice Hamiltonian as a sum of the lattice and spin parts,
\begin{equation}
\mathcal{H}_{sl} = \mathcal{H}_l + \mathcal{H}_s
\end{equation}
where the spin part $\mathcal{H}_s=\mathcal{H}_s(\mathbf{R},\mathbf{S})$ depends on atomic coordinates $\mathbf{R}$ and atomic spin vectors $\mathbf{S}=\{\mathbf{S}_i\}$. Since we use various functional forms to represent the Hamiltonians, in what follows we discuss the choice of the specific functional forms adopted in simulations.

First, we apply the umbrella sampling. We define an auxiliary Hamiltonian,
\begin{equation}
\mathcal{H}_{HO,s}=\mathcal{H}_{HO} + \mathcal{H}_s,
\end{equation}
which is a sum of the harmonic oscillators Hamiltonian and the spin Hamiltonian. The difference between this Hamiltonian and the exact Hamiltonian is
\begin{eqnarray}
\delta \mathcal{H}_{l} &=&  \mathcal{H}_{HO,s}-\mathcal{H}_{sl}\\
&=& \sum_i \left(\frac{1}{2}m\omega^2\mathbf{x}_i + C\right) - U(\mathbf{R})
\end{eqnarray}
Notably, this expression is exactly the same as that investigated in the connection with the pure MD analysis. Sampling can again be performed using Langevin thermostat simulations. Since we now also need to take into account magnetic fluctuations, the full set of equations now includes equations of motion for the spins\cite{MaPRB2012}:
\begin{equation}
\frac{d\mathbf{S}_i}{dt} = \frac{1}{\hbar}\left[\mathbf{S}_i\times\mathbf{H}_i\right]+\gamma_s \mathbf{H}_i + \bm{\xi}_i
\end{equation}
where the damping constant $\gamma_s$ and the fluctuation spin force $\bm{\xi}_i$ are related by the fluctuation-dissipation theorem $\langle \xi_{\alpha i}(t)\xi_{\beta j}(t') \rangle = 2k_B T \gamma_s \delta_{\alpha \beta} \delta_{ij}\delta (t-t')$. The effective exchange field acting on spin $i$ is
\begin{equation}
\mathbf{H}_i=-\frac{\partial \mathcal{H}_{HO,s}}{\partial \mathbf{S}_i}.
\end{equation}
Forces in (\ref{eq22}) now depend on the orientation of atomic spins
\begin{equation}
\mathbf{F}_i=  -\frac{\partial \mathcal{H}_{HO,s}}{\partial \mathbf{R}_i}.
\end{equation}

Similarly to the MD case, we evaluate the difference between the free energies of an equilibrium configuration defined by the spin-lattice Hamiltonian, and a configuration defined by the auxiliary Hamiltonian
\begin{eqnarray}
\delta F_l &=& F_{HO,s} - F_{sl}\\
&=& k_B T \ln \langle \exp (\beta\delta \mathcal{H}_{l}) \rangle_{HO,s}.
\end{eqnarray}

As the second step, we perform thermodynamic integration. We define a reference Hamiltonian
\begin{equation}
\mathcal{H}_{HO,LO}=\mathcal{H}_{HO} + \mathcal{H}_{LO},
\end{equation}
which is a sum of the harmonic oscillators Hamiltonian and the Landau Hamiltonian. The difference between the reference and the auxiliary Hamiltonians is
\begin{eqnarray}
\delta \mathcal{H}_{s} &=&  \mathcal{H}_{HO,LO}-\mathcal{H}_{HO,s}\\
&=&  \mathcal{H}_{LO}-\mathcal{H}_{s}.
\end{eqnarray}
The Hamiltonian required for carrying out thermodynamic integration can be written as
\begin{equation}
\mathcal{H}_{ti}(\lambda) = \mathcal{H}_{HO} + (1-\lambda)\mathcal{H}_{s} + \lambda\mathcal{H}_{LO}.
\end{equation}
Langevin equations of motion remain unchanged, but the effective field and the force now depend on the integration parameter $\lambda$, namely
\begin{eqnarray}
\mathbf{H}_i &=&-\frac{\partial \mathcal{H}_{ti}(\lambda)}{\partial \mathbf{S}_i},\\
\mathbf{F}_i &=&-\frac{\partial \mathcal{H}_{ti}(\lambda)}{\partial \mathbf{R}_i}.
\end{eqnarray}
The free energy difference between the equilibrium states defined by the auxiliary and reference Hamiltonians is
\begin{eqnarray}
\delta F_s &=& F_{HO,LO} - F_{HO,s}\\
&=& \int_0^1 \langle\delta\mathcal{H}_{s}\rangle_\lambda d\lambda.
\end{eqnarray}

Combining the results derived using the umbrella sampling and thermodynamic integration, we find the free energy of the equilibrium configuration defined by the spin-lattice Hamiltonian
\begin{equation}
F_{sl} = F_{HO,LO} - \delta F_l - \delta F_s.
\end{equation}
We note that this free energy $F_{sl}$ can be represented as a sum
\begin{equation}
F_{sl} = F_l + F_s,
\end{equation}
where
\begin{eqnarray}
F_l &=& F_{HO} - \delta F_l\nonumber \\
F_s &=& F_{LO} - \delta F_s.
\end{eqnarray}
The above expression has a clear meaning since $F_l$ represents a part of the free energy associated primarily with lattice excitations, whereas $F_s$ is a part of the free energy derived primarily from spin fluctuations. Since the spin and lattice degrees of freedom are coupled, and we sample through an auxiliary step, it would be inaccurate to interpret $F_l$ and $F_s$ as independent contributions from the lattice and spin subsystems. However, the two quantities still provide some qualitative insight into the relative magnitude of contributions by the two coupled subsystems to the total free energy.

\section{Simulations using literature parameterizations}
Using the methods described above, we performed MD and SLD simulations, using parameters taken from literature. All the simulations were performed using our MD and SLD program SPILADY\cite{MaCPC2016}. We used cubic simulation cells containing 16000 atoms in the bcc case and 16384 atoms in the fcc case. We explored temperatures in the range from 1K to 1400K. For each temperature, we simulated samples at nine different volumes close to an assumed equilibrium volume. A third-order polynomial was then fitted to the calculated free energies. The equilibrium volume was determined from the minimum of the polynomial. All the quantities that we describe below are interpolations corresponding to the equilibrium volume. The guessed equilibrium volume itself was computed using the same method, starting from a larger interval of trial volumes.

\begin{figure}
\centering
\includegraphics[width=8cm]{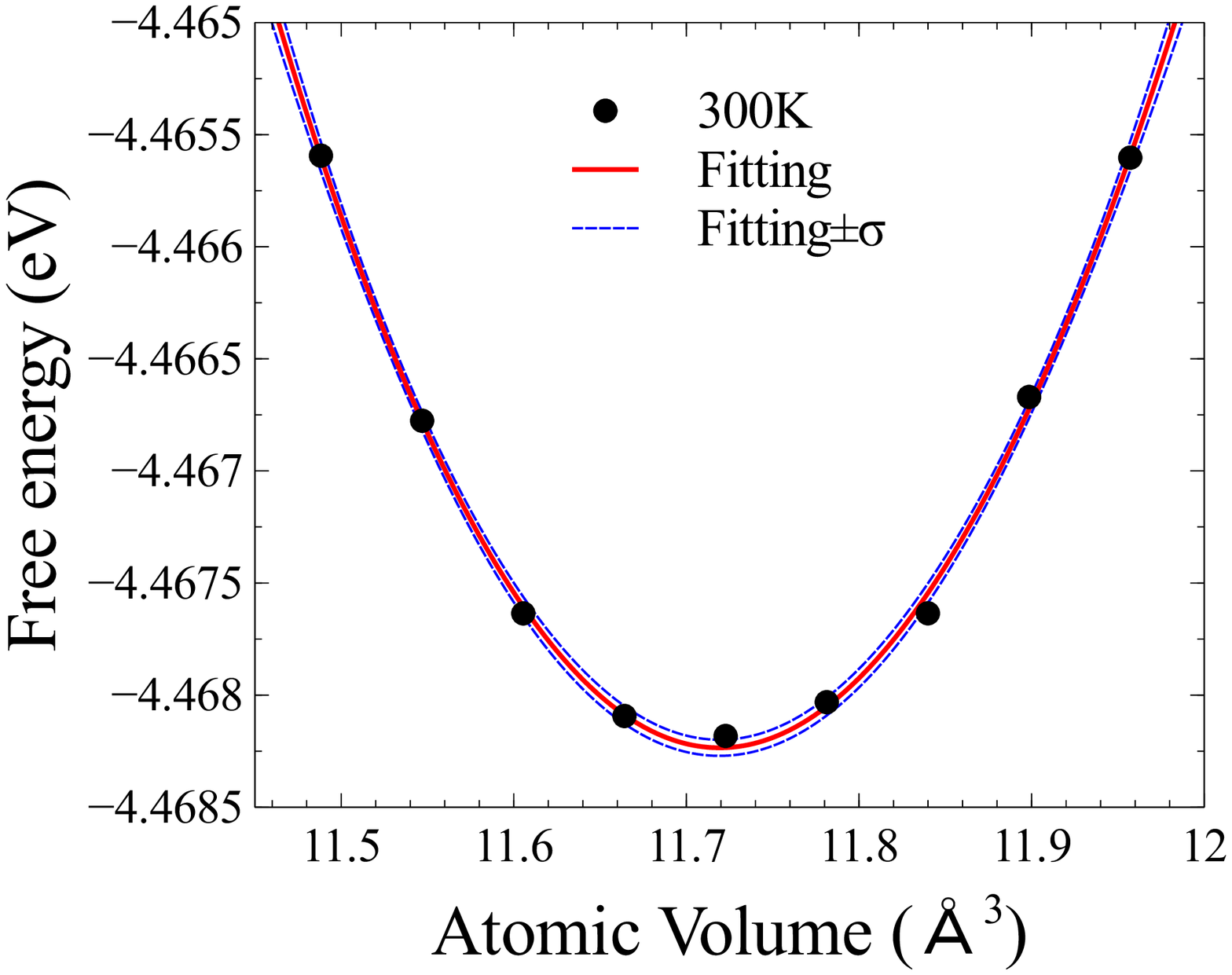}
\includegraphics[width=8cm]{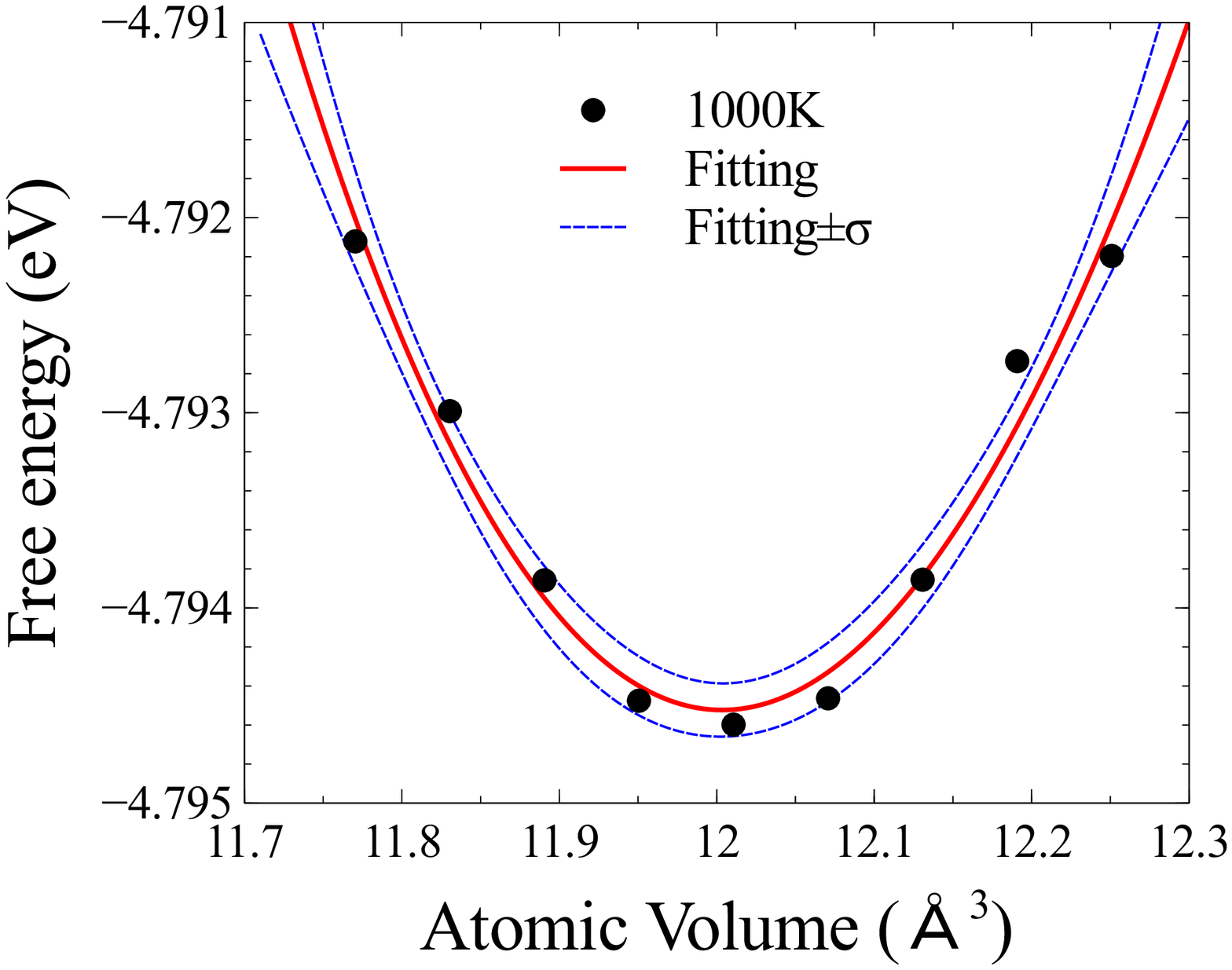}
\caption{(Color online) Free energies of bcc iron computed using spin-lattice dynamics for $T=300$K and $T=1000$K as a function of volume. The fitted curve shown in red is a third-order polynomial. The dotted curves shown in blue illustrate the standard deviation of the fitted curve shown in red. }
\label{fig1}
\end{figure}

In Fig. \ref{fig1} the free energy computed using SLD simulations is plotted as a function of volume of bcc crystal structure for 300K and 1000K. The polynomial fit interpolates the data points fairly well, although fluctuations are larger at higher temperatures, affecting the accuracy of evaluation of the equilibrium lattice constant. The dotted curves shown in blue indicate the standard deviation of the fitted curve shown in red. It is evaluated using the covariance matrix of the coefficients of the polynomial. The free energy minimum remains accurate at the sub-meV level, as illustrated by the scale of the y-axis.

All the simulation cells were thermalized to equilibrium before sampling. In MD, we take 100,000 data points when performing the umbrella sampling. In SLD, we take 200,000 data points for umbrella sampling, and 0.2ns as the total switching time for the adiabatic switching thermodynamic integration. The Marinica iron potential\cite{MalerbaJNM2010,MarinicaPRL2012} was used for both MD and SLD simulations. SLD simulations are based on the spin Hamiltonian of the form \cite{MaPRB2008,MaPRB2012}:
\begin{eqnarray}
\mathcal{H}_s &=& -\frac{1}{2}\sum_{i,j} J_{ij}(R_{ij})\left(\textbf{S}_i\cdot\textbf{S}_j - |\textbf{S}_i||\textbf{S}_j|\right)\nonumber\\
&& + \sum_i \left(A_i \textbf{S}_i^2 + B_i \textbf{S}_i^4\right)
\end{eqnarray}
where $J_{ij}$ is the exchange coupling function, and $A_i$ and $B_i$ are the Landau coefficients for atom $i$. The form of $H_s$ guarantees that the energy difference between bcc and fcc structures at 0K is the same as in the non-magnetic MD potential case.

We assume that $J_{ij}$ is a pairwise function that depends only on the distance between atoms $i$ and $j$. It has the form $J_{ij}(r) = J_0(1-r/r_c)^3\Theta(r_c-r)$, where $J_0=0.92$eV and $r_c=3.75$\AA. The value of $J_0$ is slightly larger than the one that we derived in Ref. \onlinecite{MaPRB2008} to match the experimental value of the Curie temperature $T_C$. Values of parameters $A_i=-0.744824$ eV and $B_i=0.345295$ eV are taken from Ref. \onlinecite{MaPRB2012}. We note that the ground state of this Hamiltonian is ferromagnetic regardless of whether the underlying crystal structure is bcc or fcc.

\begin{figure}
\centering
\includegraphics[width=8cm]{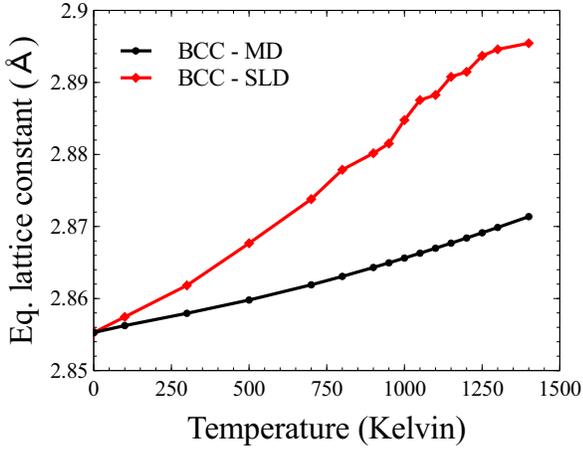}
\includegraphics[width=8cm]{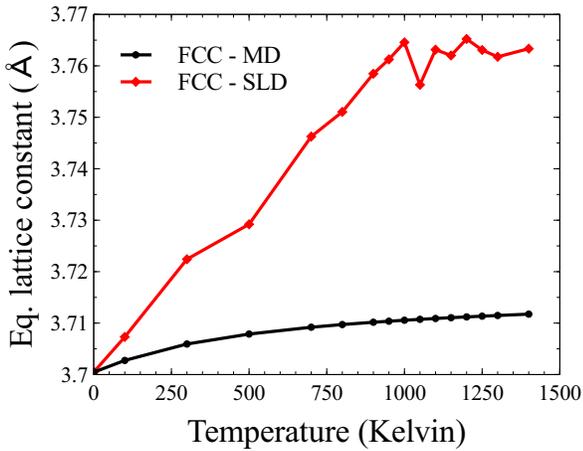}
\caption{(Color online) Equilibrium lattice constants of bcc and fcc phases as functions of temperature, computed using MD and SLD simulations.}
\label{fig2}
\end{figure}

\begin{figure}
\centering
\includegraphics[width=8cm]{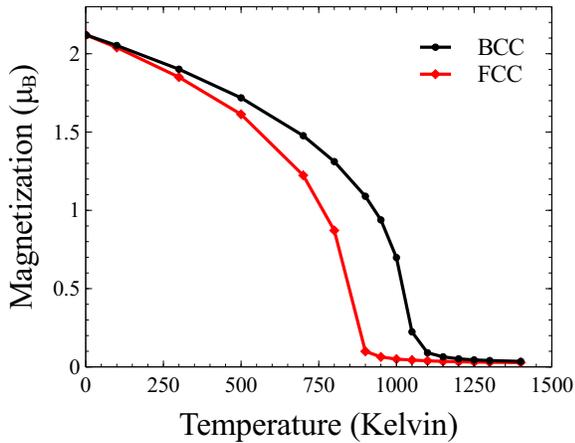}
\caption{(Color online) Magnetization as a function of temperature, computed using SLD simulations.}
\label{fig3}
\end{figure}

Fig. \ref{fig2} shows equilibrium lattice constants as functions of temperature, predicted by MD and SLD simulations. Magnetic excitations enhance thermal expansion of both bcc and fcc structures. The curves derived from SLD simulations flatten in the vicinity of $T_C$ (Fig. \ref{fig3}). Fluctuation of the curves result primarily from polynomial fitting.

Since we use the same spin Hamiltonian for bcc and fcc cases, they both adopt ferromagnetic ground states at temperatures below $T_C$. However, experimental data for fcc iron indicate that it has a relatively low N\'eel temperature $T_N$ of 67K\cite{GonserJAP1963,JohansonPRB1970}. Although the precise nature of magnetic configuration at temperatures below $T_N$ is debatable, the net magnetization is zero. This differs from our simulations, and we will address the issue in the following sections.

\begin{figure}
\centering
\includegraphics[width=8cm]{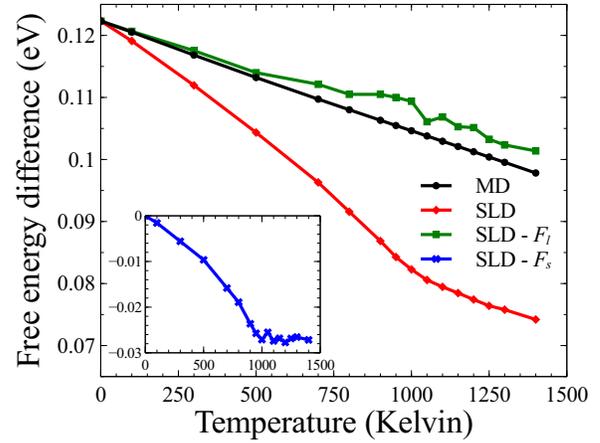}
\includegraphics[width=8cm]{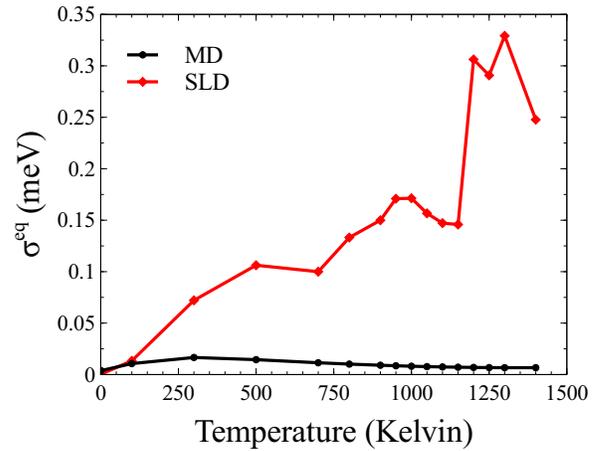}
\caption{(Color online) (a) Difference between free energies of fcc and bcc phases plotted as a function of temperature. Data derived from MD and SLD are shown, with lattice and spin contributions shown separately. (b) Standard deviation of the free energy at the equilibrium lattice constant.}
\label{fig4}
\end{figure}

In Fig. \ref{fig4}, we plotted the free energy difference between fcc and bcc phases as a function of temperature. We define $\Delta F^{fcc-bcc} = F^{fcc} - F^{bcc}$ in such a way that when this value is positive, bcc phase is stable and vice versa. In MD, $\Delta F^{fcc-bcc}$ is always positive and does not approach zero when temperature increases. Interestingly, if we include magnetic excitations, then $\Delta F^{fcc-bcc}$ decreases significantly. We also show $\Delta F^{fcc-bcc}_l = F^{fcc}_l - F^{bcc}_l$ on the same graph, and $\Delta F^{fcc-bcc}_s = F^{fcc}_s - F^{bcc}_s$ in the inset. The free energy difference predicted by SLD simulations largely originates from the spin subsystem, though it is not sufficient to stabilize the fcc phase at high temperature. We also note that the derivative of $\Delta F^{fcc-bcc}_s$ with respect to temperature is small near the Curie temperature $T_C$ of the bcc phase. Standard deviations of the calculated free energies are also shown, which are all in the sub-meV level.

Although we find that the magnitude of the magnetic part of the free energy is significant, we see that parameters taken from literature have no chance of success in predicting the bcc-fcc phase transition. The main deficiency of existing parameterizations is that the effective interatomic potentials were all fitted without considering magnetic excitations. Even if we add a spin part to the Hamiltonian in an {\it ad-hoc} manner, this still does not fully account for the free energy contribution from the spin system. A better approach to deriving parameters for spin-lattice dynamics simulations is necessary to model magnetic iron on the atomic scale.

\section{Parametrization}
In what follows, we present a new derivation of parameters for spin-lattice dynamics simulations of bcc and fcc iron. We start by fitting a non-magnetic iron potential, and augment it by the Heisenberg-Landau Hamiltonian.

\subsection{Non-magnetic iron potential}
We fitted a non-magnetic iron potential using an interatomic potential fitting program \textit{potfit}\cite{BrommerMSMSE2015,BrommerMSMSE2007,BrommerPM2006}. It fits a many-body potential to a user-defined functional form. Parameters of the potential are fitted using the force matching method\cite{ErcolessiEL1994}, using the total energy and forces taken from \textit{ab initio} data. All of our \textit{ab initio} calculations were performed using VASP\cite{KressePRB1993,KressePRB1994,KresseCMS1996,KressePRB1996}. We use the GGA-PBE\cite{PerdewPRL1996,PerdewPRL1997} pseudo-potential with 14 valence electrons. The plane wave energy cutoff is 450 eV.

We first generate {\it ab initio} data for the non-magnetic iron. The structures include perfect bcc and fcc lattices, and simulation cells with distortions such as rhombohedral and tetragonal shape, at various volumes. We also produced {\it ab initio} data for amorphous structures and structures containing defects. The functional form and parameters of the fitted potential are given in Appendix. Since we are only interested in the energy and free energy differences between bcc and fcc structures, we are not going to discuss other features of our non-magnetic iron potential here.

\begin{figure}
\centering
\includegraphics[width=8cm]{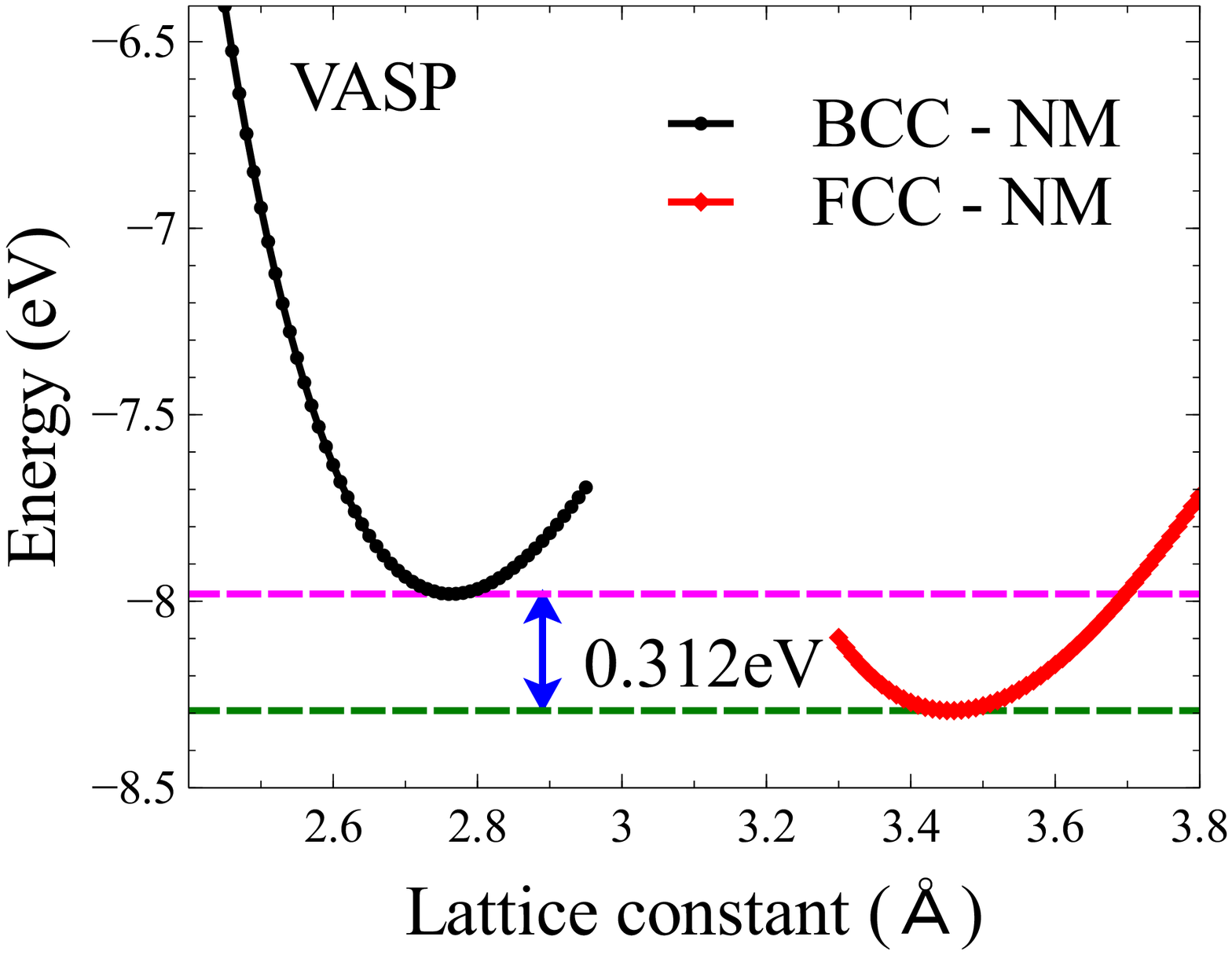}
\includegraphics[width=8cm]{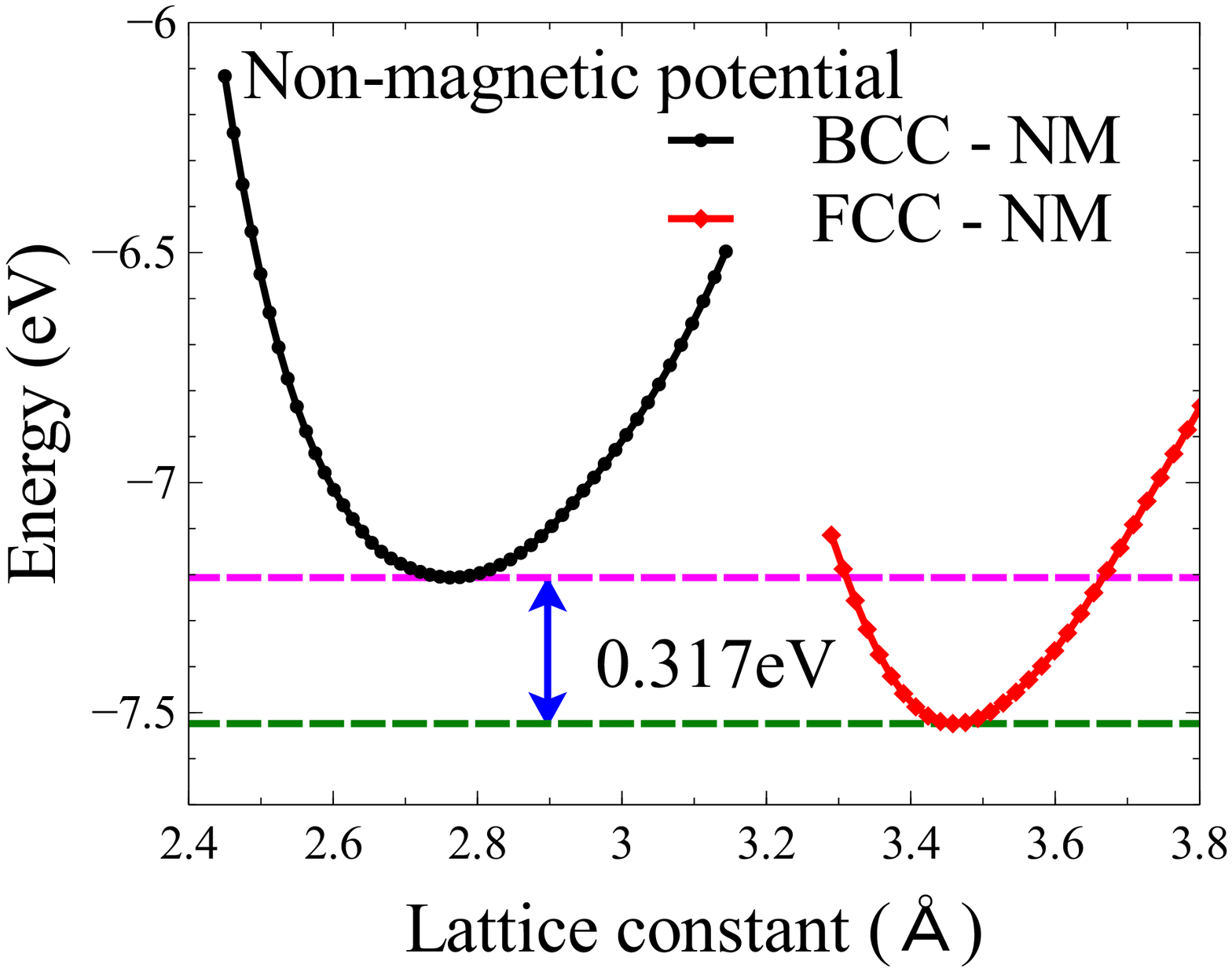}
\caption{(Color online) Energy per atom for non-magnetic bcc and fcc structures calculated using VASP and the fitted non-magnetic interatomic potential.}
\label{fig5}
\end{figure}

Fig. \ref{fig5} shows \textit{ab initio} energies of non-magnetic bcc and fcc phases at various lattice constants. The minimum energy of the fcc phase is 0.312eV lower than the minimum energy of the bcc phase, in agreement with data from Ref. \onlinecite{HerperPRB1999}. Fcc structure is more stable when magnetism is not taken into account. The curves computed using non-magnetic potential appear similar, and the energy difference is 0.317eV. The difference between the absolute values of \textit{ab initio} data and non-magnetic potential data is due to the different choice of the reference points. Since we take the cutoff distance of the potential as 5.3\AA, the energies of all the \textit{ab initio} data points are reduced by the energy of a perfect bcc structure with lattice constant $a=5.3\times2/\sqrt{3}=6.1199$\AA, where the nearest neighbour distance is 5.3\AA. Interatomic forces remain unaffected by this procedure.

\subsection{Magnetic contributions}
\begin{figure}
\centering
\includegraphics[width=8cm]{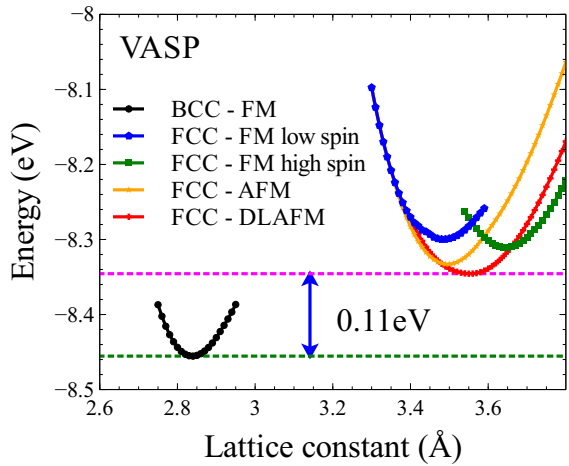}
\caption{(Color online) Energies of ferromagnetic bcc structure, and ferromagnetic (high spin and low spin), single layer anti-ferromagnetic, and double layers anti-ferromagnetic fcc structures calculated using VASP.}
\label{fig6}
\end{figure}

Fig. \ref{fig6} shows VASP data for the energy of magnetic bcc and fcc phases at various lattice constants. In the bcc case, we only show the data for the FM collinear ground state. In the fcc case, there are a number of magnetic configurations that all have comparable energies. We show {\it ab initio} data for the FM, AFM, and double layer AFM (DLAFM) magnetic configurations. If we impose a constraint and consider only the collinear magnetic configurations, the DLAFM state has the lowest energy. This also agrees with \textit{ab initio} results given in Ref. \onlinecite{HerperPRB1999}.

The energy of the FM bcc phase is now 0.11eV lower than that of the DLAFM fcc phase. We now need to find a way of describing magnetic excitations using a magnetic Hamiltonian added to the non-magnetic Hamiltonian. This magnetic Hamiltonian should also describe interactions between magnetic moments that are not explicitly evident from the VASP data.

There are various ways to describe interactions between magnetic moments. One can use the spin spiral method\cite{OkatovPRB2011} or spin-cluster expansion method\cite{SingerPRL2011}. We choose the spin Hamiltonian in the Heisenberg-Landau form\cite{LavrentievCMS2010,LavrentievPRB2010,MaPRB2012},
\begin{eqnarray}
\mathcal{H}_s &=& -\frac{1}{2}\sum_{i,j} J_{ij}(R_{ij})\mathbf{M}_i\cdot\mathbf{M}_j\nonumber \\
&&+ \sum_i\left(A(\rho_i)\mathbf{M}_i^2 + B(\rho_i)\mathbf{M}_i^4\right)\label{eq46}
\end{eqnarray}
where $\mathbf{M}=-g\mu_B\mathbf{S}_i$ is the magnetic moment of atom $i$, $g=2.0023$ is the electron $g-$factor, $\mu_B$ is the Bohr magneton, $A(\rho_i)$ and $B(\rho_i)$ are the Landau coefficients that depend on the effective electron density $\rho_i$. This is the same $\rho_i$ that enters the non-magnetic interatomic potential.

\begin{figure}
\centering
\includegraphics[width=8cm]{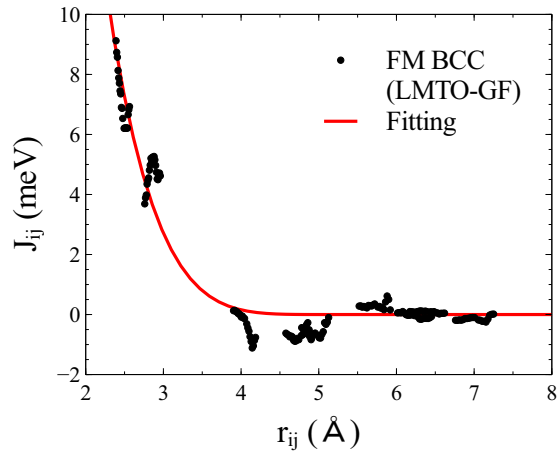}
\includegraphics[width=8cm]{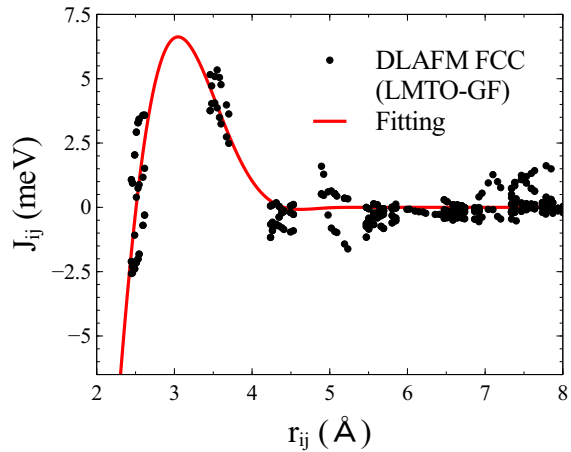}
\caption{(Color online) Exchange coupling $J_{ij}$ for the ferromagnetic bcc and double layers anti-ferromagnetic fcc structures calculated using the LMTO-GF method, and the fitting functions.}
\label{fig7}
\end{figure}

The first step is to calculate values of $J_{ij}$ from the lowest energy state of bcc and fcc phases. The exchange coupling functions are calculated using \textit{ab initio} electronic structure multiple-scattering formalism. We use the method and program developed by van Schilfgaarde {\it et al.} \cite{LiechtensteinJMMM1987,SchilfgaardeJAP1999}. It is based on the linear muffin-tin orbital approximation combined with Green's function technique (LMTO-GF). We calculated values of parameters $J_{ij}$ involving various neighbours over a range of variation of the lattice constant. Fig. \ref{fig7} shows the calculated values and the fitted curves.  \textit{Ab initio} data for the bcc case are smoother, whereas the data for the fcc cases are more scattered. A possible reason is that magnetic configuration of collinear FM state in the bcc phase is fundamentally simpler than the DLAFM state of the fcc phase. To match the data, we used different functional forms for the bcc and fcc cases. The functional forms and values of parameters are given in Appendix.

\begin{figure}
\centering
\includegraphics[width=8cm]{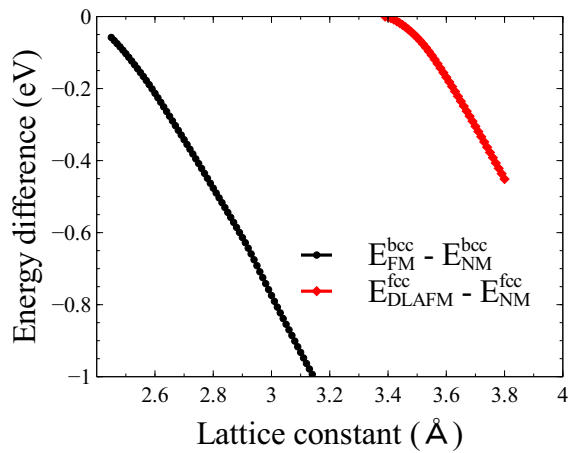}
\caption{(Color online) Difference between energies of magnetic and non-magnetic states of ferromagnetic bcc and double layers antiferromangetic fcc structures.}
\label{fig8}
\end{figure}

\begin{figure}
\centering
\includegraphics[width=8cm]{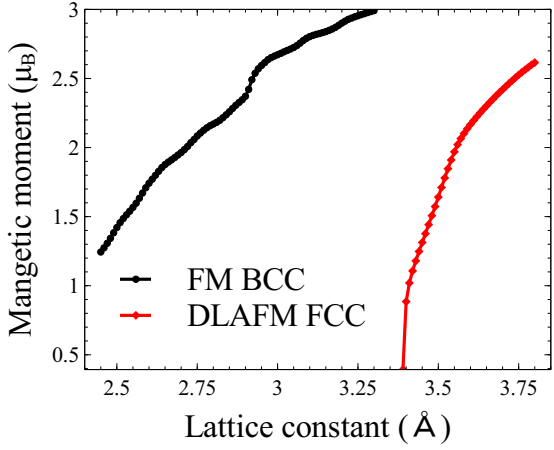}
\caption{(Color online) Magnitude of magnetic moments computed for ferromagnetic bcc and double layers antiferromangetic fcc structures.}
\label{fig9}
\end{figure}

We now evaluate the Landau coefficients. We define a temporary Hamiltonian with no magnetic moment interactions, assuming that on-site magnetic moments can be treated as order parameters, i.e.
\begin{equation}
\mathcal{H}'_s = \sum_i\left(A'\mathbf{M}_i^2 + B'\mathbf{M}_i^4\right).
\label{eq47}
\end{equation}
Since we know the difference between energies of magnetic and non-magnetic configurations (Fig. \ref{fig8}), and also the magnitude of magnetic moments on each atoms (Fig. \ref{fig9}) as functions of lattice constant, we can identify the energy difference per atom and the magnitude of magnetic moment as
\begin{eqnarray}
\Delta E &=& A'M_0^2+B'M_0^4\\
M_0 &=& \sqrt{-A'/2B'}\ne 0,
\end{eqnarray}
or $M_0=0$ if the non-magnetic state is more stable. We obtain values of $A'$ and $B'$ at various lattice constants by solving the above equations.

\begin{figure}
\centering
\includegraphics[width=8cm]{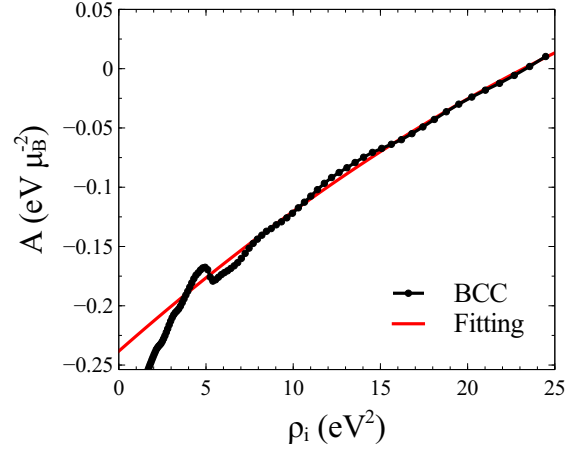}
\includegraphics[width=8cm]{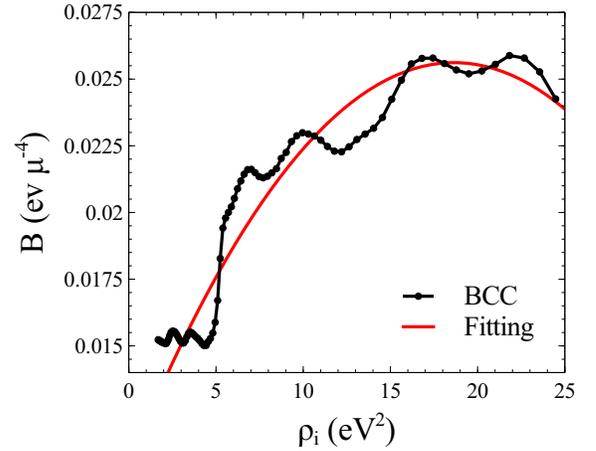}
\caption{(Color online) Landau coefficients $A$ and $B$ as functions of the effective electron density $\rho$ for bcc structures. }
\label{fig10}
\end{figure}

\begin{figure}
\centering
\includegraphics[width=8cm]{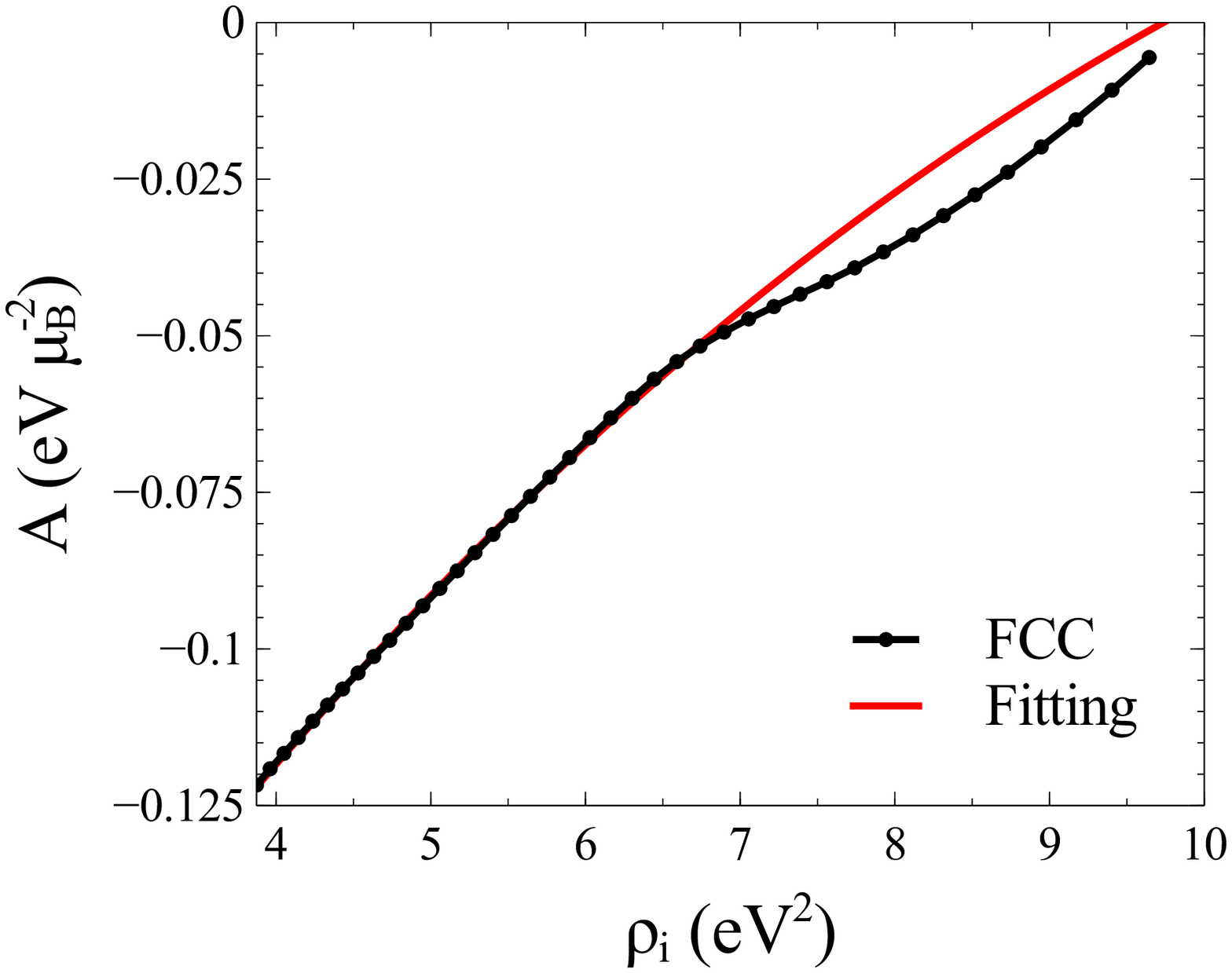}
\includegraphics[width=8cm]{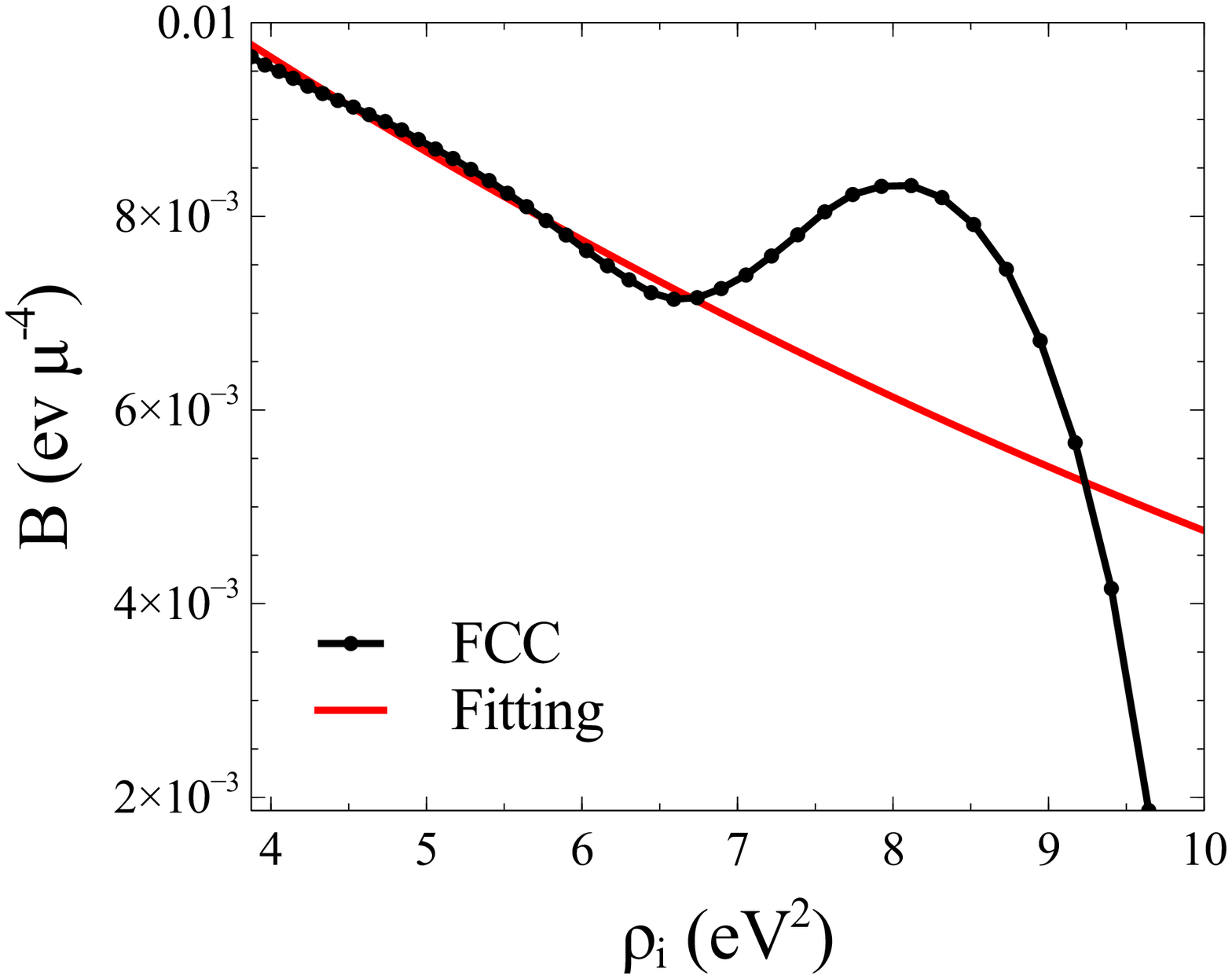}
\caption{(Color online) Landau coefficients $A$ and $B$ as functions of effective electron density $\rho$ for fcc structures. }
\label{fig11}
\end{figure}

We now need to relate the values of the Landau parameters to the values of parameters characterizing the Hamiltonian that describes interacting magnetic moments. We equate Eq. \ref{eq46} and \ref{eq47}, and find
\begin{eqnarray}
A \mathbf{M}_i^2 &=& A' \mathbf{M}_i^2+ \frac{1}{2}\sum_j J_{ij}\mathbf{M}_i\cdot\mathbf{M}_j\\
B &=& B'.
\end{eqnarray}
Using the fitted function $J_{ij}$ and considering magnetic configurations of perfect crystals, we find the values of Landau coefficients $A$ and $B$. These coefficients are plotted as functions of the effective electron density $\rho_i$ for perfect crystals (Fig. \ref{fig10} and \ref{fig11}). Again, we use different functional forms for fitting results for bcc and fcc cases, to match various features of the curves. The functional forms and numerical parameters are given in Appendix. The strong scatter of values for $B(\rho_i)$ is due to the small value of $M_0$ corresponding to the small size of the simulation box.

Using the above procedure, we generated several sets of parameters, which were derived using different methods and have various functional forms. The parameters that have been selected are those that match experimental results well. Although all of the parameters produce qualitatively similar predictions, Landau coefficients may need to be adjusted through the choice of fitting intervals to achieve sub-meV accuracy of free energy calculations. In all cases, \textit{ab initio} data provide the foundation for the fitting procedure.

\section{Structural Phase Transitions}
We calculated the free energies of bcc and fcc phases using the above new sets of parameters. We performed both MD and SLD simulations. In MD, we used the non-magnetic potential, and performed umbrella sampling calculations with 100,000 data points. In SLD, we used the non-magnetic potential with Heisenberg-Landau Hamiltonian, and used the two step approach in the free energy calculations. We took 300,000 data points for umbrella sampling, and 0.2ns as the total switching time in the adiabatic switching thermodynamic integration. The magnetic configuration of bcc and fcc cases are initialized as FM and DLAFM states, respectively. We explored a large temperature range from $1\times 10^{-5}$K to 2000K. All that samples are thermalized to equilibrium before sampling.

\begin{figure}
\centering
\includegraphics[width=8cm]{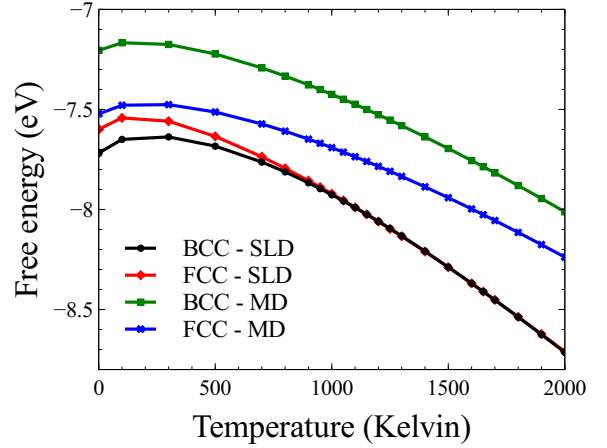}
\caption{(Color online) Free energy of bcc and fcc phases as functions of temperature. Both MD and SLD results are shown.}
\label{fig12}
\end{figure}

\begin{figure}
\centering
\includegraphics[width=8cm]{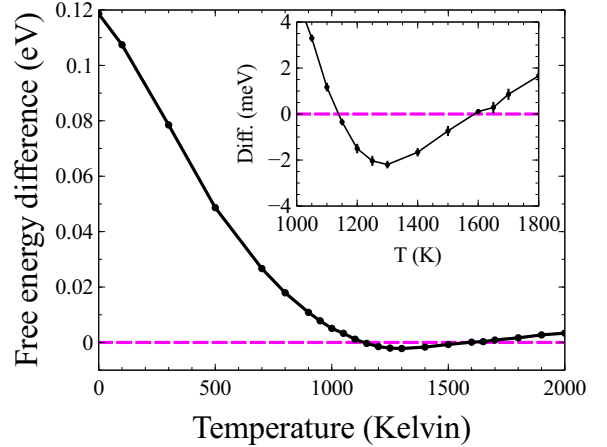}

\caption{(Color online) Difference between free energies of bcc and fcc phases plotted as a function of temperature, where both magnetic excitations and lattice vibrations are included. Calculations are performed using a non-magnetic potential combined with the Heisenberg-Landau Hamiltonian. The inset shows a magnified part of the same figure.  }
\label{fig13}
\end{figure}

\begin{figure}
\centering
\includegraphics[width=8cm]{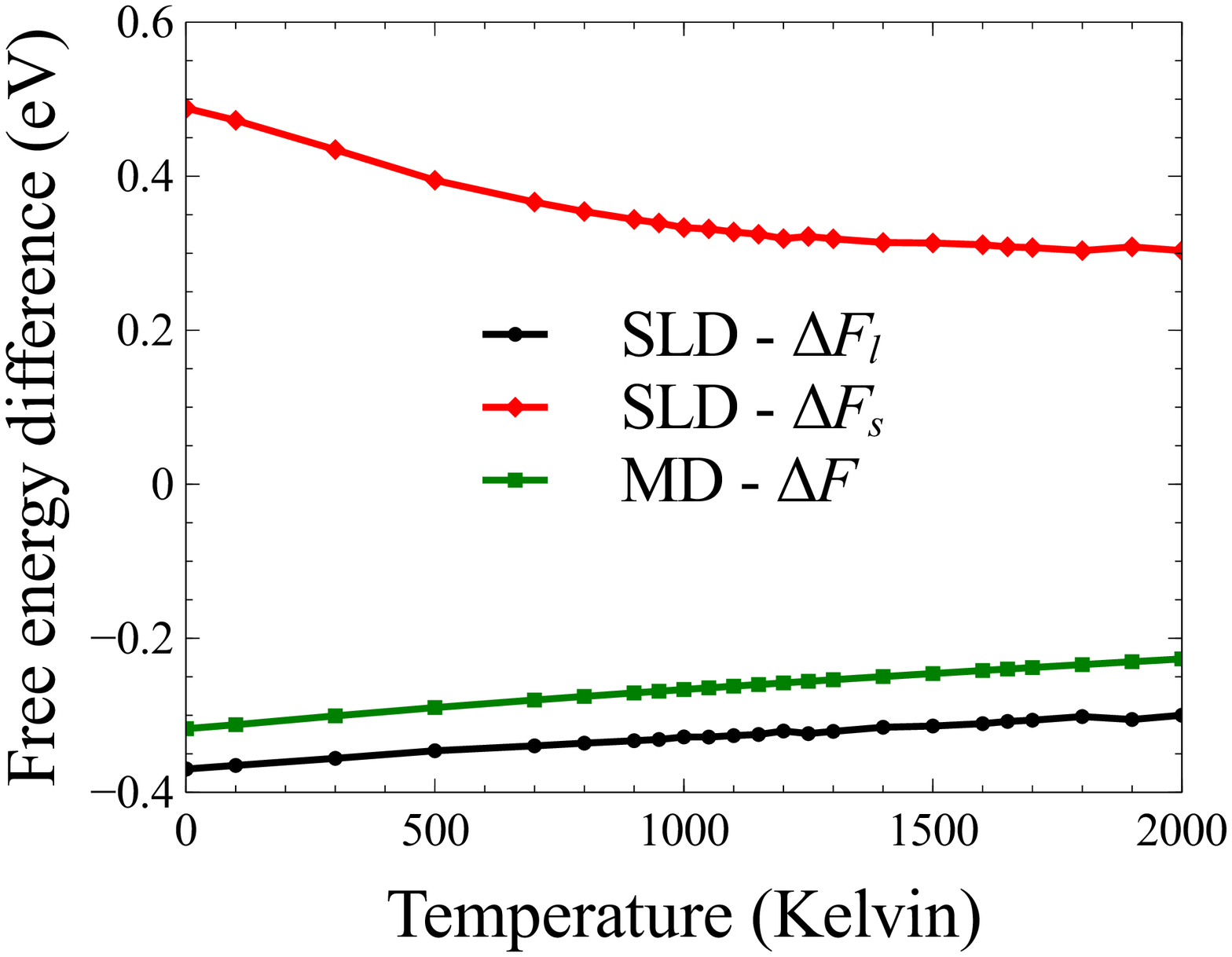}
\includegraphics[width=8cm]{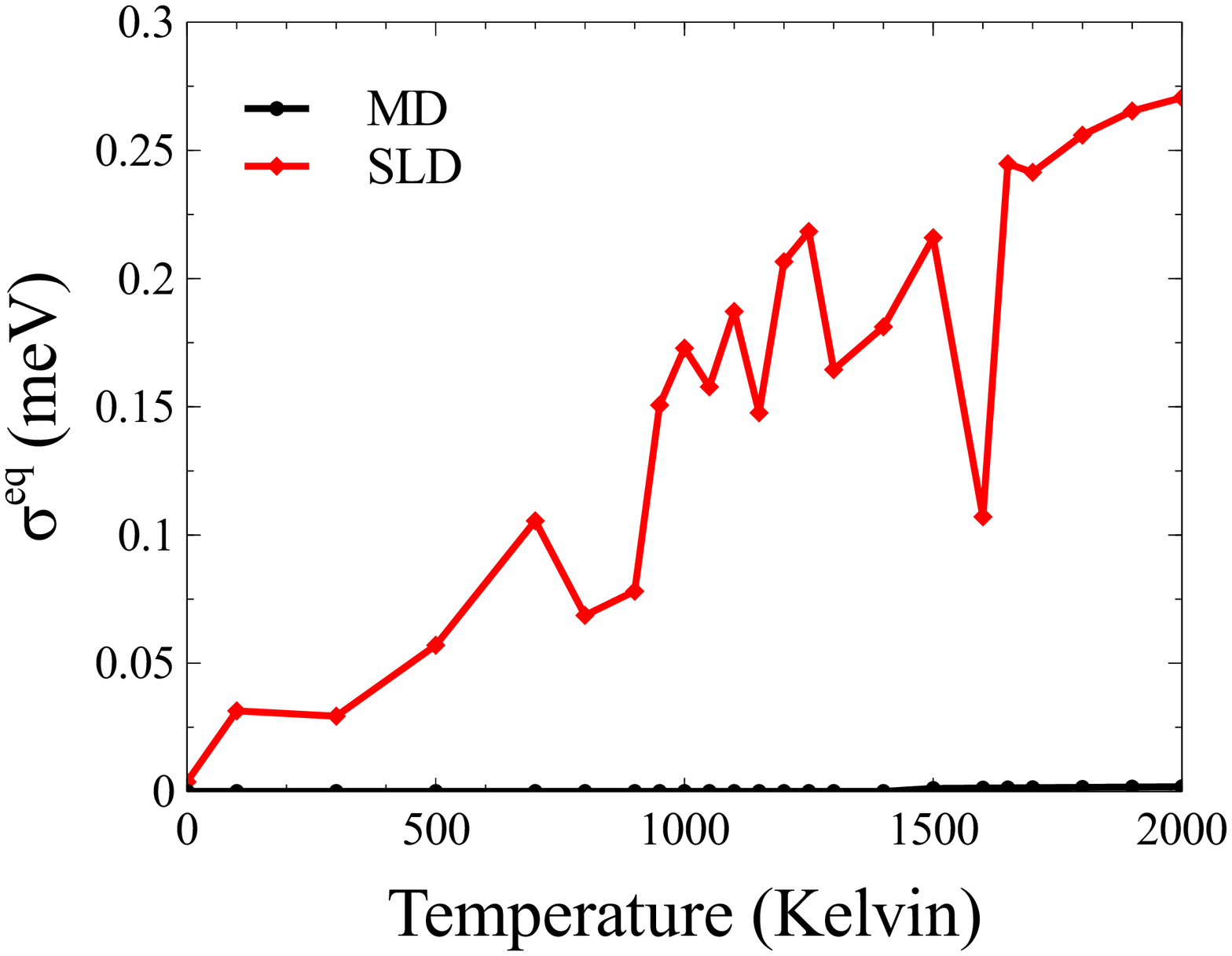}
\caption{(Color online) (a) Lattice and spin contributions to the free energy difference between bcc and fcc structures. (b) The standard deviation of the free energy at the equilibrium lattice constant. }
\label{fig14}
\end{figure}

The most significant results of this paper are illustrated in Fig. \ref{fig12} and \ref{fig13}. Fig. \ref{fig12} shows the calculated free energies of bcc and fcc phases at the equilibrium volume. In the MD case, the free energy of the fcc phase is always lower than the free energy of the bcc phase. In SLD, the bcc phase has lower free energy initially, but the curves corresponding to bcc and fcc phases approach each other at higher temperature. There are two intersections between the curves, which can be seen if one follows the difference $\Delta F^{fcc-bcc}$ plotted in Fig. \ref{fig13}. The curve crosses the zero line at around 1130K and 1600K. These temperatures are close to the experimentally observed values of $\alpha$-$\gamma$ and $\gamma$-$\delta$ phase transitions at $T_{\alpha-\gamma}$=1185K and $T_{\gamma-\delta}$=1667K, respectively. The free energy difference at the minimum is close to 2meV. This value agrees with the MCE\cite{LavrentievCMS2010,LavrentievPRB2010} and RPA\cite{KormannCOSSMS2016} results, and shows that $\alpha-\gamma-\delta$ transitions are associated with fairly small free energy differences between the competing phases, of the order of 1 meV.

The reason for the difference between MD and SLD simulations may be understood by considering free energy contributions from magnetic excitations and lattice vibrations. Fig. \ref{fig14} shows plots of $\Delta F^{fcc-bcc}_l$ and $\Delta F^{fcc-bcc}_s$ as functions of temperature. For comparison, we also plotted $\Delta F^{fcc-bcc}_{MD}$ calculated using the non-magnetic potential. The value of $\Delta F^{fcc-bcc}_l$ is similar to $\Delta F^{fcc-bcc}_{MD}$. Both of them increase relatively linear as functions of temperature. On the other hand, $\Delta F^{fcc-bcc}_s$ decreases initially but flattens out near the Curie temperature $T_C$ of the bcc phase. Hence the change of $\Delta F^{fcc-bcc}_s$ is mainly due to the difference in the degree of disorder associated with magnetic configurations involved. If the temperature is higher than $T_C$, the bcc phase is in the paramagnetic state, where the long range magnetic order vanishes. Magnetic configurations of fcc and bcc phases become similar, making the entropy difference smaller. This interpretation is consistent with recent experimental findings\cite{NeuhausPRB2014} on phonon dispersion of iron. Experimental data suggest that the $\gamma$ phase forms as a result of interplay between electronic and vibrational contributions to entropy, whereas the $\delta$ phase is due primarily to the contribution of vibrational entropy. The two crossing points result from the interplay between free energy contributions derived from lattice and spin excitations. The standard deviation of the free energy remains in the sub-meV level.

\begin{figure}
\centering
\includegraphics[width=8cm]{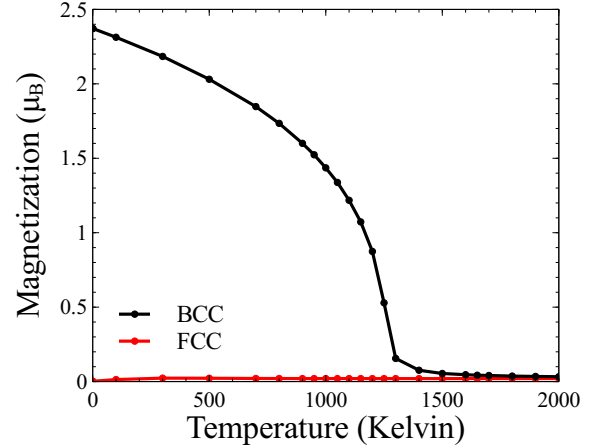}
\caption{(Color online) Magnetization as a function of temperature computed using SLD simulations and the non-magnetic potential supplemented with the Heisenberg-Landau Hamiltonian.}
\label{fig15}
\end{figure}

\begin{figure}
\centering
\includegraphics[width=8cm]{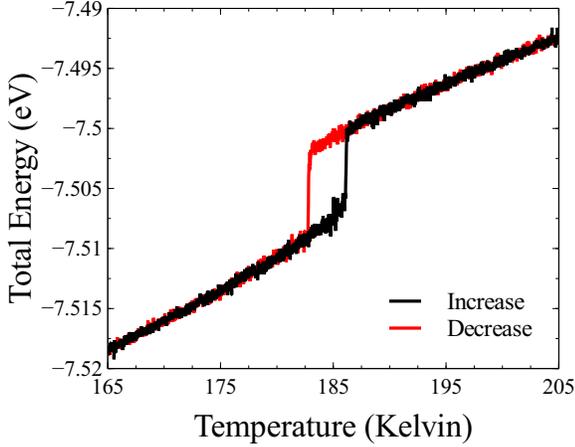}
\includegraphics[width=8cm]{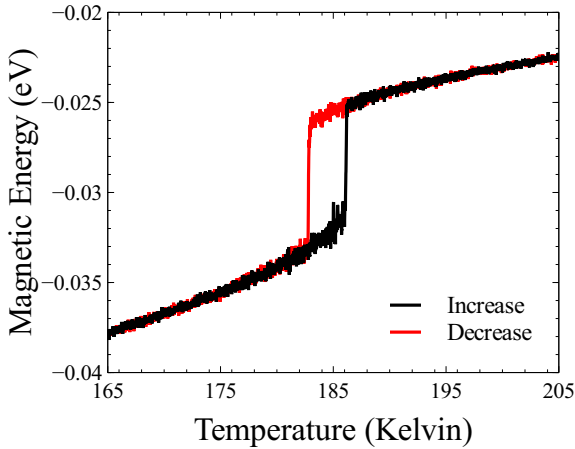}
\caption{(Color online) The total energy and magnetic energy per atom as functions of temperature calculated by gradually increasing and decreasing the temperature of the thermostat.}
\label{fig16}
\end{figure}

\begin{figure}
\centering
\includegraphics[width=8cm]{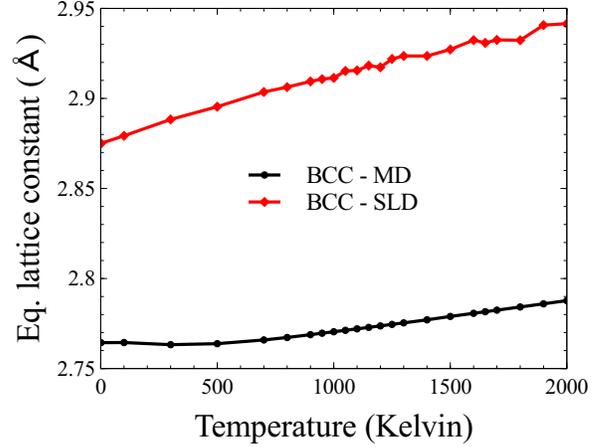}
\includegraphics[width=8cm]{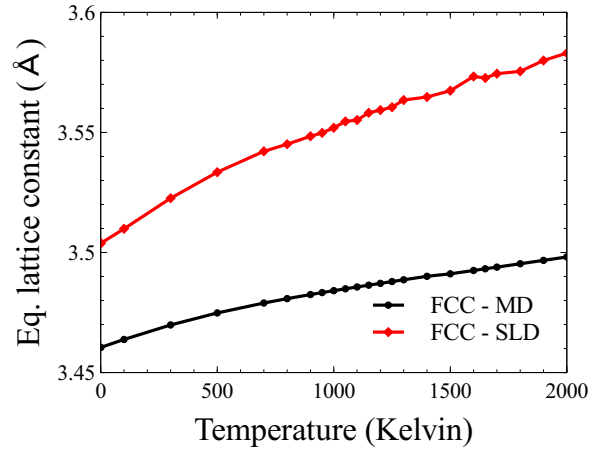}
\caption{(Color online) Equilibrium lattice constants of bcc and fcc phases as functions of temperature, computed using MD and SLD simulations using new potentials and parameters.}
\label{fig17}
\end{figure}

The magnetization curve of iron is plotted in Fig. \ref{fig15}. For the bcc phase, the existing model predicts a relatively high Curie temperature at around 1300K. For the fcc case, the predicted magnetization is zero. We investigated the N\'eel temperature $T_N$ of the fcc phase by  changing the value of energy per atom as a function of temperature (Fig. \ref{fig16}). We thermalized two simulation cells to 165K and 205K. Then, we gradually increased the temperature of the simulation cell equilibrated at 165K to 205K. This was performed by increasing the thermostat temperature linearly over the time interval of 4 ns. Then, we decrease the temperature of the simulation cell equilibrated at 205K to 165K. We filtered the output data by averaging over every 1000 data points. We see that the magnetic subsystem undergoes a first order transition at around 185K. A jump of about 0.01 eV in the magnetic energy can be observed. The change of the total energy is fully accounted for by the magnetic energy. The calculated $T_N$ is higher than the experimental value of 67K. However, such value is obtained from experiments on small particles\cite{GonserJAP1963,JohansonPRB1970} where the role of local stresses is unclear. Besides, the temperatures of the $\alpha-\gamma$ and $\gamma-\delta$ phase transitions are significantly higher than $T_N$.

We also show equilibrium lattice constants of bcc and fcc phases predicted by MD and SLD simulations in Fig. \ref{fig17}. The addition of the spin Hamiltonian changes the value of the equilibrium lattice constant even at 0K. Fundamentally, this agrees with \textit{ab initio} data shown in Fig. \ref{fig5} and \ref{fig6}. The spin subsystem of the material affects mechanic properties through its contribution to the total free energy.

\section{Conclusions}
Starting from a large amount of \textit{ab initio} data, we fitted non-magnetic many-body potentials and Heisenberg-Landau Hamiltonians for bcc and fcc iron. We performed free energy calculations using umbrella sampling and thermodynamics integration. The free energy has been sampled by molecular dynamics and spin-lattice dynamics simulations. Our method provides a reasonably consistent way of assessing the phase stability of magnetic iron within a unified dynamic picture. It treats both magnetic excitations and lattice vibrations and their coupling self-consistently. The bcc-fcc ($\alpha$-$\gamma$) and fcc-bcc ($\gamma$-$\delta$) phase transitions in magnetic iron are reproduced using  newly fitted potential and parameters. The structural phase stability of magnetic iron is governed by non-collinear magnetic excitations and lattice vibrations, in agreement with other experimental and theoretical results. The maximum free energy difference between bcc and fcc phases is about 2 meV.

\appendix
\section{Non-magnetic iron potential}

\begingroup
\begin{table}
\begin{ruledtabular}
\begin{tabular}{ | l | r | }
 $\phi$ & $-4.483075702293698016e-04$ \\
\hline
 $r_0^t$ & $ 2.0000000000000000e+00$ \\
 $r_1^t$ & $ 2.2000000000000000e+00$ \\
 $r_2^t$ & $ 2.6000000000000000e+00$ \\
 $r_3^t$ & $ 3.2000000000000000e+00$ \\
 $r_4^t$ & $ 3.8000000000000000e+00$ \\
 $r_5^t$ & $ 4.6000000000000000e+00$ \\
 $r_6^t$ & $ 5.3000000000000000e+00$ \\
 $t_0$ & $ 2.5999782982854347e+00$ \\
 $t_1$ & $ 2.9319480072508499e+00$ \\
 $t_2$ & $ -2.8388905185188360e+00$ \\
 $t_3$ & $ -1.0267419494754382e-01$ \\
 $t_4$ & $ 1.5484736035888333e-02$ \\
 $t_5$ & $ -7.2805743511785065e-02$ \\
 $t_6$ & $ -3.6343523861565924e-03$ \\
\hline
 $r_0^V$ & $ 2.3254531341916498e+00$ \\
 $r_1^V$ & $ 2.3889005055990276e+00$ \\
 $r_2^V$ & $ 2.5614990650026459e+00$ \\
 $r_3^V$ & $ 2.5615004425308658e+00$ \\
 $r_4^V$ & $ 2.8344513929093051e+00$ \\
 $r_5^V$ & $ 2.8321879787700808e+00$ \\
 $r_6^V$ & $ 2.6382534884695783e+00$ \\
 $r_7^V$ & $ 3.4262631740080707e+00$ \\
 $r_8^V$ & $ 3.8479639767860356e+00$ \\
 $r_9^V$ & $ 3.8515517885908994e+00$ \\
 $r_{10}^V$ & $ 4.3740210397021579e+00$ \\
 $r_{11}^V$ & $ 4.4054035197078845e+00$ \\
 $r_{12}^V$ & $ 4.5503412747697087e+00$ \\
 $r_{13}^V$ & $ 4.7731075757035732e+00$ \\
 $r_{14}^V$ & $ 5.3000000000000000e+00$ \\
 $V_0$ & $ 2.2831054190426084e+01$ \\
 $V_1$ & $ -2.1062362139531867e+01$ \\
 $V_2$ & $ 5.6190823955741749e+00$ \\
 $V_3$ & $ 8.0795758060570382e+00$ \\
 $V_4$ & $ -8.5213153270399573e+01$ \\
 $V_5$ & $ 9.0355710040623180e+01$ \\
 $V_6$ & $ -8.3613137262443793e+00$ \\
 $V_7$ & $ -3.4250845501053456e-01$ \\
 $V_8$ & $ 5.2035042290453923e+01$ \\
 $V_9$ & $ -5.1583785613198948e+01$ \\
 $V_{10}$ & $ 4.0569674844835752e+00$ \\
 $V_{11}$ & $ -5.0779874829818361e+00$ \\
 $V_{12}$ & $ 1.7323802861372730e+00$ \\
 $V_{13}$ & $ -3.5971267571846299e-01$ \\
 $V_{14}$ & $ -1.1478647839739256e-01$ \\
\end{tabular}
\end{ruledtabular}
\caption{Parameters of the non-magnetic iron potential.}
\label{table1}
\end{table}
\endgroup

The functional form of the interatomic potential broadly follows the conventional embedded atom method (EAM) representation
\begin{equation}
U\left(\mathbf{R}_1, \mathbf{R}_2, ... \right)=\sum_i F(\rho_i) + \frac{1}{2}\sum_{i,j}V_{ij}\left(R_{ij}\right),
\end{equation}
where $\mathbf{R}_i$ is the position of atom $i$, $\rho_i$ is the effective electron density and $V_{ij}$ is a pairwise function that depends only on the distance between atoms $i$ and $j$. The many-body part of the potential takes the same form as that proposed by Mendelev \textit{et al.}\cite{MendelevPM2003} and Ackland \textit{et al.}\cite{AcklandJPCM2004}
\begin{equation}
F(\rho_i) = -\sqrt{\rho_i}+\phi\rho_i^2
\end{equation}
where $\phi$ is a parameter. The effective electron density $\rho_i$ is defined in a slightly different way from the conventional EAM potential. We write
\begin{equation}
\rho_i = \sum_j t_{ij}^2
\end{equation}
where $t_{ij}=t_{ij}(R_{ij})$ is a pairwise hopping integral, which we take a function of the distance between the atoms $R_{ij}$. We note that the derivative of $\rho_i$ with respect to $R_{ij}$ is
\begin{equation}
\frac{\partial \rho_i}{\partial R_{ij}}=2t_{ij}\frac{\partial t_{ij}}{\partial R_{ij}}.
\end{equation}
$t_{ij}$ and $V_{ij}$ are given by the third-order splines
\begin{eqnarray}
t_{ij}(x) &=& \sum_n t_n (r_n^t - x)^3 \Theta(r_n^t-x)\\
V_{ij}(x) &=& \sum_n V_n (r_n^V - x)^3 \Theta(r_n^V-x)
\end{eqnarray}
where $n$ are knots, $t_n$, $V_n$ are parameters with dimensionality eV\AA$^{-3}$, and $r_n^t$ and $r_n^V$ are given in \AA $\;$ units. Their values are given in Table \ref{table1}.

\section{Exchange coupling and Landau coefficients}

\begingroup
\begin{table}
\begin{ruledtabular}
\begin{tabular}{ | l | r | }
 bcc & \\
\hline
$J_0$	&	$	1.7613094778950000e-01	$\\
$r_{c	ut}$	& $	5.3000000000000000e+00	$\\
$a_0$	&	$	-2.3827723674043900e-01	$\\
$a_1$	&	$	1.2945703172205700e-02	$\\
$a_2$	&	$	-1.1518969922985000e-04	$\\
$b_0$	&	$	1.0600315078586900e-02	$\\
$b_1$	&	$	1.6104913287021000e-03	$\\
$b_2$	&	$	-4.3178188078544200e-05	$\\
\hline
 fcc & \\
\hline
$J_0$	&	$	1.1095507874951400e-01	$	\\
$r_{cut}$	&	$	5.3000000000000000e+00	$	\\
$b$	&	$	1.6502332463388100e+00	$\\	
$c$	&	$	-4.1373722623161200e+00	$	\\
$a_0$	&	$	3.1803486683085200e-01	$\\	
$a_1$	&	$	6.0141682907976200e-02	$\\	
$\rho_a$	&	$	2.2852502987397700e+01	$\\	
$b_0$	&	$	1.4290243674270400e-02	$\\	
$b_1$	&	$	0.0000000000000000e+00	$\\	
$\rho_b$	&	$	3.2563330708156800e+01	$\\	
\end{tabular}
\end{ruledtabular}
\caption{Parameters for exchange coupling and Landau coefficients}
\label{table2}
\end{table}
\endgroup

The exchange coupling function and Landau coefficients, expressed as functions of electron density, have different functional forms for bcc and fcc cases. In the bcc case, we use the following form
\begin{eqnarray}
J_{ij}(r_{ij}) &=& J_0(1-r_{ij}/r_{cut})^5,\\
A(\rho_i) &=& a_0 + a_1 \rho_i + a_2 \rho_i^2,\\
B(\rho_i) &=& b_0 + b_1 \rho_i + b_2 \rho_i^2.
\end{eqnarray}
In the fcc case, we take
\begin{eqnarray}
J_{ij}(r_{ij}) &=& J_0\sin(br_{ij}+c)(1-r_{ij}/r_{cut})^3,\\
A(\rho_i) &=& a_0(1-\rho_i/\rho_a)^3 + a_1,\\
B(\rho_i) &=& b_0(1-\rho_i/\rho_b)^3 + b_1.
\end{eqnarray}
The units of $J_{ij}$, $A$ and $B$ are eV$\mu_B^{-2}$, eV$\mu_B^{-2}$ and eV$\mu_B^{-4}$, respectively. The cutoff distance $r_{cut}$ is in \AA $\;$ units. All the parameters are listed in Table \ref{table2}.

\begin{acknowledgments}
This work has been carried out within the framework of the EUROfusion Consortium and has received funding from the Euratom research and training programme 2014-2018 under grant agreement No 633053 and from the RCUK Energy Programme [grant number EP/P012450/1]. To obtain further information on the data and models underlying this paper please contact PublicationsManager@ccfe.ac.uk*. The views and opinions expressed herein do not necessarily reflect those of the European Commission. We also acknowledge EUROFusion for the provision of Marconi supercomputer facility at CINECA in Italy. The authors are grateful to M.-C. Marinica for stimulating discussions.
\end{acknowledgments}

\end{document}